\documentclass[twocolumn,superscriptaddress,amsmath,amssymb]{revtex4-2}
\usepackage{color}
\usepackage{graphicx}% Include figure
\usepackage{subfigure}% subplot
\usepackage{booktabs}
\usepackage{dcolumn}% Align table columns on decimal point
\usepackage{bm}% bold math
\usepackage{amsmath}%more math font
\usepackage{verbatim}%multi-line "comment" environment
\usepackage{lineno}
\usepackage[T1]{fontenc}

\newcommand{\beq}{\begin{eqnarray}}
\newcommand{\eeq}{\end{eqnarray}}

\begin{document}

\title{Thermodynamic crossovers in supercritical fluids}
\author{Xinyang Li}
\affiliation{CAS Key Laboratory of Theoretical Physics, Institute of Theoretical Physics, Chinese Academy of Sciences, Beijing 100190, China}
\affiliation{School of Physical Sciences, University of Chinese Academy of Sciences, Beijing 100049, China}

\author{Yuliang Jin}
\email{yuliangjin@mail.itp.ac.cn}
\affiliation{CAS Key Laboratory of Theoretical Physics, Institute of Theoretical Physics, Chinese Academy of Sciences, Beijing 100190, China}
\affiliation{School of Physical Sciences, University of Chinese Academy of Sciences, Beijing 100049, China}
\affiliation{Wenzhou Institute, University of Chinese Academy of Sciences, Wenzhou, Zhejiang 325000, China}

\date{\today}

\begin{abstract}
Can liquid-like and gas-like states be distinguished beyond the critical point, where the liquid-gas phase transition no longer exists and conventionally only a single supercritical fluid phase is defined? Recent experiments and simulations report strong evidence of dynamical crossovers above the critical temperature and pressure~\cite{simeoni2010widom,  gorelli2013dynamics, pipich2018densification}.Despite using different criteria, {many} existing theoretical explanations consider a single   crossover line separating liquid-like and gas-like states in the supercritical fluid phase~\cite{xu2005relation, fisher1969decay, brazhkin2012two,  nishikawa1995correlation, ploetz2019gas, woodcock2013observations, strong2018percolation}.
We argue that such a single-line scenario is  inconsistent with the supercritical behavior of the Ising model, which has two crossover lines due to its symmetry,  violating the universality principle of critical phenomena. 
To reconcile the inconsistency, we define two thermodynamic crossover lines in supercritical fluids as boundaries of liquid-like, indistinguishable and gas-like states. 
Near the critical point, the two crossover lines follow critical scalings with exponents of the Ising universality class, supported by calculations of theoretical models and analyses of experimental data from the standard  database~\cite{NIST}. 
The upper line  agrees with crossovers independently estimated from  the inelastic X-ray scattering  data of supercritical argon~\cite{simeoni2010widom, gorelli2013dynamics}, and from the small-angle neutron scattering data of supercritical carbon dioxide~\cite{pipich2018densification}.
The lower line is verified by the equation of states for the compressibility factor. This work provides a fundamental framework for understanding supercritical physics in general phase transitions.
\end{abstract}

\maketitle
According to textbook knowledge, no liquid-gas phase transitions exist in the supercritical fluid state of matter.   Then, how does a liquid transform into a gas (or vice versa) along a path by passing the critical point?  Many studies  propose that there should be a line of {\it supercritical crossovers} between liquid-like and gas-like states. 
Such a crossover line divides the phase diagram above the critical point into regimes with different physical properties, which can be connected without going through any thermodynamic singularity.
The following supercritical crossover lines have been defined.

\begin{figure*}[!htbp]
  \centering
  \includegraphics[width=1\linewidth]{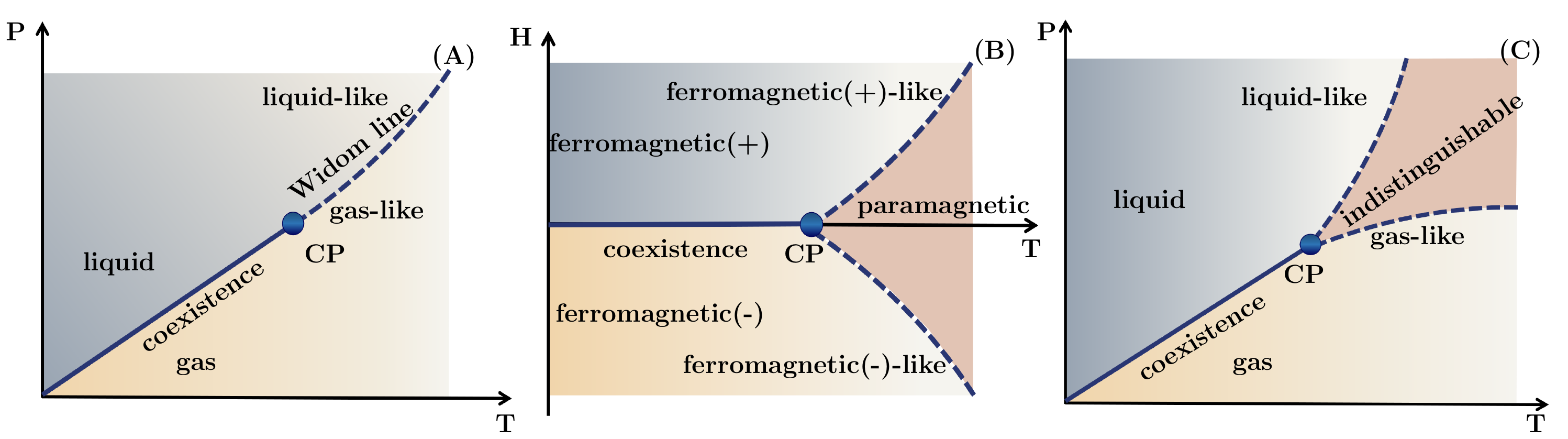}
  \caption{{\bf Schematic phase diagrams.} (A) Liquid-gas (pressure $P$ vs temperature $T$ ) phase diagram with a Widom  line (dashed) separating liquid-like and gas-like regimes above the critical point (CP).
  (B) Phase diagram of the Ising model, with two supercritical crossover lines (dashed)
  %on the positive side and negative side of external magnetic field respectively, 
  separating positive ferromagnetic-like, paramagnetic, and negative ferromagnetic-like regimes. 
  %\YJ{Change F(+)-like to ferromagnetic-like(+); same for -}
  (C) Liquid-gas phase diagram with two supercritical crossover lines (dashed), $L^{\pm}$, separating liquid-like, indistinguishable, and gas-like regimes.}
%  \YJ{move gas-like closer to the $L^-$ line. In B and C, use a different color for the indistinguishable area.}}
  %and indistinguishable regime
  %for our study, with two dynamic crossover lines similar to Ising model separating liquid-like regime, gas-like regime and indistinguishable regime. \blue{Here CP represents critical point, and F(+)/F(-) represents ferromagnetic(+)/ferromagnetic(-).}
 %  \YJ{I think we can remove B. The phase diagram of the Ising model is already in SI with exact results. }}
  \label{fig:schematic}
\end{figure*}

%scenarios can be found. 
%The notion and definition of Widom line can be straightforwardly generalized to other phase transitions, such as  liquid-liquid phase transitions~\cite{xu2005relation}.
(I) {\it Widom line}~\cite{jones1956specific, xu2005relation, brazhkin2011widom, ruppeiner2012thermodynamic, banuti2015crossing, may2012riemannian, luo2014behavior, banuti2017similarity, corradini2015widom, gallo2014widom, de2021widom} (see Fig.~\ref{fig:schematic}A), defined as the line of maxima in 
{isobaric specific heat $C_P$}
%thermodynamic response functions (e.g., specific heat or compressibility) 
under the fixed pressure $P$ (or temperature $T$) condition.
The maximum value increases approaching the critical point, and diverges at the point. 
The lines of maxima determined according to different response functions are assumed to converge into a single line near the critical point, which  emanates from the critical point.

%While thermodynamic response functions diverge at the critical point, above the critical point they display peaks instead of divergence. The Widom line, 

(II) {\it Fisher-Widom line}~\cite{fisher1969decay, de1994decay, vega1995location, dijkstra2000simulation, evans1993asymptotic, tarazona2003fisher, stopper2019decay},  defined as the boundary between states with exponential (gas-like) and oscillatory (liquid-like) long-range decays of the pair correlation function.
Liquid-like and gas-like states are distinguished based on  the structure of configuration, i.e., the spatial distribution of particles.
%This line does not pass through the critical point.
%are treated as dense gases. As the density of a system increases from the dilute gas limit, at a certain point the spatial distribution of particles should deviate from the typical behavior of gases. Explicitly, the Fisher-widom line is

(III) {\it Frenkel line}~\cite{brazhkin2012two, yoon2018two, brazhkin2013liquid, bolmatov2013thermodynamic, bolmatov2014structural, prescher2017experimental, bolmatov2015frenkel, fomin2018dynamics, proctor2019transition, fomin2015dynamical, brazhkin2012universal, cockrell2021transition}, 
%where the vibrational time matches the structural relaxation time (diffusion time).
which separates liquid-like and gas-like states based on whether
viscoelastic dynamics are present. 
Within the vibrational time, liquids behave like amorphous solids -- particles vibrate around their equilibrium positions; beyond the structural relaxation time (diffusion time), particles diffuse,  and the equilibrium configuration decorrelates from the initial one. 
As $T$ increases (or $P$ decreases), this liquid-like picture breaks down at the Frenkel line, where the vibration time becomes comparable with the structural relaxation time, and the system turns into a gas-like state.

%of equilibrium positions — a criterion defines the Frenkel line. 

(IV)  {\it Nishikawa line}~\cite{nishikawa1995correlation, nishikawa1998fluid, matsugami2014theoretical}, defined by the ``ridge'' of the density fluctuation profile on the $P$-$T$ phase diagram, where the density fluctuation maximizes. 

(V) {\it Symmetry line}~\cite{ploetz2019gas}, defined at the  zero skewness in the particle number distribution for  the grand canonical ensemble.
Liquid-like behavior is characterized by a negative skewness indicating the favor of particle deletion, whereas gas-like behavior is characterized by a positive skewness indicating the favor of particle insertion.

%, which  corresponds to a maximum in the pair particle number fluctuation density with respect to pressure along an isotherm. 

(VI)  {\it Percolation lines}~\cite{woodcock2013observations, woodcock2016thermodynamics}, determined by the loci of the percolation transition of 
 available volume accessible to any single mobile atom,
 and that of the cluster formed by bonded atoms.
 The percolation line of the hydrogen bond network has also been studied in supercritical water~\cite{strong2018percolation}, but the method cannot be applied to supercritical fluids without hydrogen bonds.

 {(VII) Boundaries of the pseudo-boiling transitional region~\cite{banuti2015crossing, maxim2021thermodynamics}. The transition upon crossing the Widom line has been referred to as supercritical pseudo-boiling~\cite{banuti2015crossing}, a phenomenon obeying similar laws as subcritical boiling, according to which the Widom line is also named as a {\it pseudo-boiling line}. At a constant pressure $P$, the pseudo-boiling transition starts at a temperature $T^{\rm -}(P)$, where the isobaric specific heat $C_P$ starts to deviate from its liquid value, and ends at $T^{\rm +}(P)$, when $C_P$ approaches its ideal gas value. 
 The two lines, $T^{\rm \pm}(P)$, are {\it pseudo-boiling boundaries}.
 }

 %Pseudo-boiling lines $T^{\pm}$  ~\cite{banuti2015crossing, maxim2021thermodynamics}, determined by the loci where the isobaric specific heat capacity approaches the liquid heat capacity ($T^-$) and ideal gas heat capacity ($T^+$). It is considered that the supercritical phase transition takes place over this finite temperature interval.}

%\red{It is well known that Ising model is a model for a ferromagnetic phase transition, but that is not all. It is also a model for describing liquid-gas transition when mapping onto the lattice gas model~\cite{PhysRev.87.410, friedli_velenik_2017}. From the point of view of Landau theory, they share the same language. Moreover,} 
The liquid-gas system and the Ising model both exhibit a critical point. 
The liquid-gas transition is not associated with the change of structural order, while the magnetic phase transition in the Ising model is of the order-disorder type.
Despite this difference, the liquid-gas critical point and the magnetic phase transition in the three-dimensional (3D) Ising model belong to the same $O(1)$ universality class~\cite{kadanoff1976scaling, fisher1983scaling}.
According to the principle of universality in statistical mechanics, they should display critical scalings with the same exponents. {In Ref.~\cite{luo2014behavior}, a general linear scaling theory and molecular dynamics simulations are employed to obtain the Widom line for different kinds of critical points, including the liquid-gas critical point where the slope of the coexistence line is positive in the $P$-$T$ plane (see Fig.~\ref{fig:schematic}A for the supercritical branch of the Widom line), the liquid-liquid critical point where the slope is negative, and the magnetic phase transition in the Ising model where the slope is zero (see Fig.~\ref{fig:schematic}B)~\cite{luo2014behavior}. However, the behavior of the Widom line is non-universal in the three cases.
In general, the above proposals (I-VII) of crossover lines in supercritical fluids are all} theoretically inconsistent with the supercritical crossover lines of the Ising model. To show that, let us first discuss the analogies between the two systems, and the supercritical behavior of the Ising model. 

%\blue{From the phase transition point of view, the ferromagnetic(+) - ferromagnetic(-) first order phase transition in the Ising model corresponds to the liquid-gas phase first order phase transition, in which case both of their order parameters are not zero, so they belong to order-order transition. The ferromagnetic-paramagnetic critical phase transition in the Ising model corresponds to the liquid/gas-indistinguishable critical phase transition in liquid-gas systems, where both of two order parameters change from non-zero to zero, therefore they are in the order-disorder type. From the point of view of physical quantities,}

The pressure $P$ and susceptibility
$\kappa_T \equiv  \frac{P_{\rm c}}{\rho_{\rm c}}  \left(\frac{\partial \rho}{\partial  P} \right)_T$ (or the
compressibility $\beta_T = \frac{1}{\rho} \left( \frac{\partial \rho}{\partial P} \right)_{T}$)  in {liquid-gas systems}
are analogous of the magnetic field $H$ and magnetic susceptibility $\chi = \left( \frac{\partial m}{\partial H} \right)_{T}$ in the 
{Ising model}, where
$P_{\rm c}$ and $\rho_{\rm c}$ are the critical pressure and density. Here $P$ and $H$ play the role of external fields;  $\kappa_T$ and $\chi$ characterize the response of  the order parameter (i.e., the liquid-gas density difference $\delta \rho $ and the  magnetization $m$ respectively) to the external field.
%While the criteria of Fisher-Widom line and  Frenkel line are not applicable, those of other crossover lines can be generalized for the Ising model. 
Following the definition of Widom line, it is immediately apparent that in the Ising model, the maxima of $\chi$ at constant $H$ form two crossover lines (denoted by  $L^{+}$ and $L^{-}$) above the critical temperature $T_{\rm c}$. Due to the $Z_2$ symmetry of the Ising model, the two lines $L^{\pm}$  are perfectly symmetric with respect to the coexistence line $H=0$. 
%\YJ{In SI, $h$ is used, unify $H$ and $h$}
Along $L^{\pm}$ lines, the field and order parameter follow critical scalings, 
%The scalings of $L^{\pm}$, 
\beq
|H^{\pm}| \sim (T - T_{\rm c})^{\beta+\gamma},
\label{eq:scalingH}
\eeq
and
\beq
|m^{\pm}| \sim (T - T_{\rm c})^{\beta},
\label{eq:scalingm}
\eeq
established by analytic calculations of the mean-field model and Monte-Carlo numerical simulations of 2D and 3D models (see Supplementary Information SI Sec.~S1), where $\beta$ and $\gamma$ are universal critical exponents (for the 3D Ising universality class, $\beta \simeq 0.3265$ and $\gamma \simeq 1.237$~\cite{guida1998critical}). 

%(see also Supplementary Information (SI) Table S1).

\begin{figure*}[!htbp]
  \centering
  \includegraphics[width=0.8\linewidth]{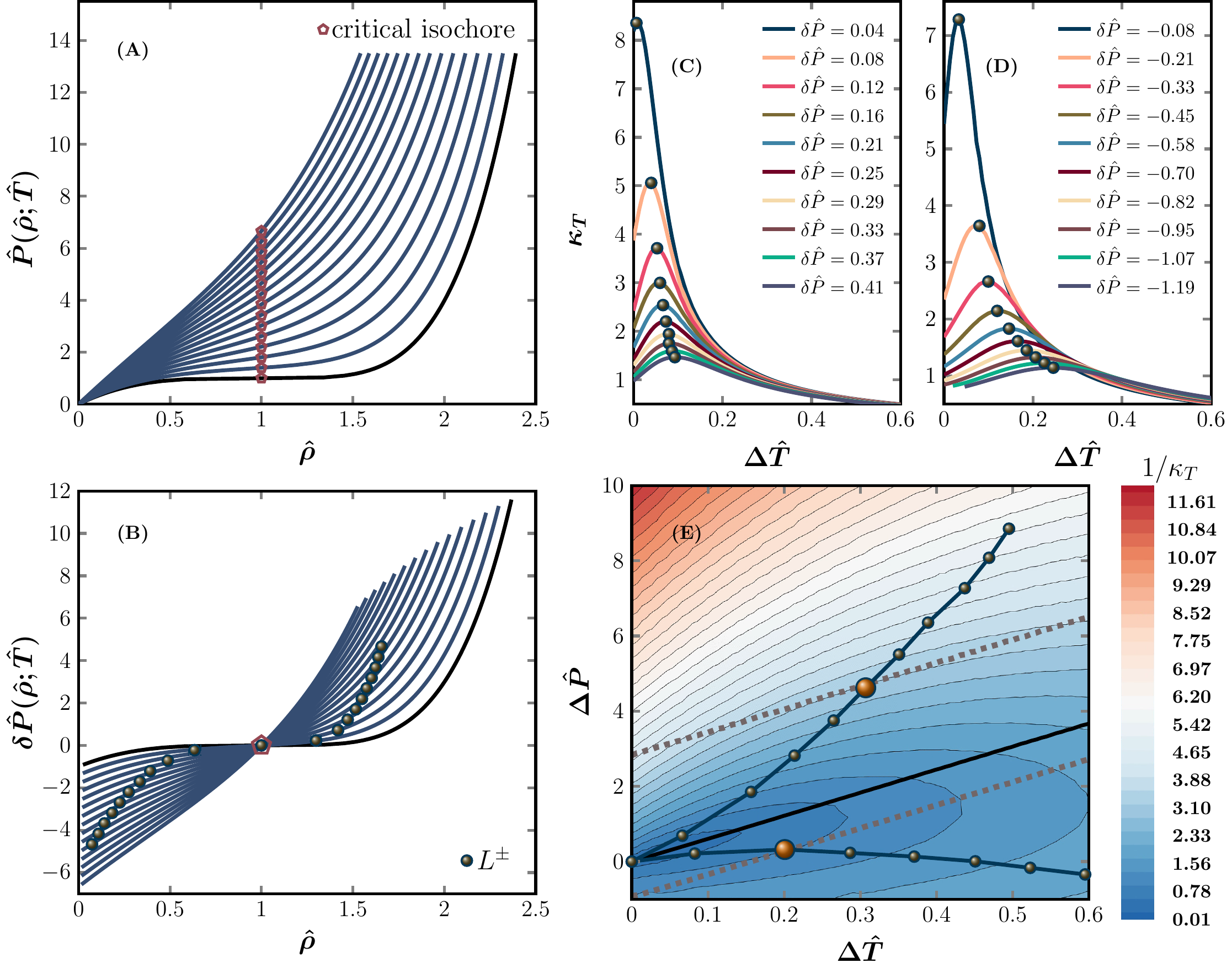}
  \caption{{\bf Determination of supercritical crossover lines $L^{\pm}$ in argon}.
  (A) Isothermal EOSs, $\hat{P}(\hat{\rho}; \hat{T})$, for $\hat{T} =$ 1.00, 1.07, 1.13, 1.20, 1.27, 1.33, 1.40, 1.46, 1.53, 1.60, 1.66, 1.73, 1.80, 1.86, 1.93.
  (from bottom to top).
  %\YJ{The critical isochore in the legend is wrong.}
 % above the critical temperature. The lowest (black)  line is the critical isotherm, and the other (dark blue) lines are other isotherms. 
 % \YJ{Indicate critical isochore in the legend}
 % The points on the critical isochore are marked by brown pentagons. 
 (B) Shifted EOSs $\delta \hat{P}(\hat{\rho}; \hat{T})$.
 % where $\delta \hat{P}(\hat{\rho}; \hat{T}) = \hat{P}(\hat{\rho}; \hat{T}) -  \hat{P}(1; \hat{T})$.
 % \YJ{indicate $L^{\pm}$ in the legend of $B$.}
%  for which the pressure $\hat{P}(1; \hat{T})$ along the   critical isochore is subtracted from each EOS. 
(C-D) Susceptibility 
 $\kappa_T$ 
    %= \left( \frac{\partial \hat{\rho}}{\partial  \hat{P}} \right)_{\hat{T}}$ is plotted 
    as a function of $\Delta \hat{T}$, for a few fixed $\delta \hat{P}$.
    %(C) $\delta \hat{P}>0$ and (D) $\delta \hat{P}<0$; 
    The loci of peaks determine (C) $L^+$ and (D) $L^-$ lines, for $\delta \hat{P}>0$ and $\delta \hat{P}<0$  respectively. 
    %\YJ{change $\kappa$ to $\kappa_T$}
    (E) The color map of  $1/\kappa_T$ in the $\Delta\hat{P}-\Delta\hat{T}$ diagram, with constant $1/\kappa_T$ contour lines (solid thin  lines). 
    As an example, we show two parallel paths (dashed brown lines)  to  the critical isochore (think black line). 
     Such a parallel path has a point of tangency  (big orange ball) with a certain constant $1/\kappa_T$ contour; along this parallel path,  the point of tangency corresponds to the maximum of 
     $\kappa_{T}$ as defined in $L^{\pm}$.
    %is indicated by the big yellow circle. 
    In panels (B-E), the $L^{\pm}$ lines are represented by the same symbol.}
   % \YJ{Add $1/\kappa_T$ to the colorbar.}}
    %The meaning of symbols are the same in all panels. 
 % \YJ{Mark critical isochore in (A) and (B). Mark $L^{\pm}$ in (C) and (D). Add (D) and change (D) to (E). C-D): Difficult to see the curves; maybe reduce the x-axis range; indicate $\delta P$.  Use same symbol in all panels and figures. For example, use the same symbol for $L^{\pm}$ in (B) and (E). Increase the width of lines whenever possible. Increase the size of points whenever possible. Add the colorbar in (E)}}
  \label{fig:method}
\end{figure*}

The  $L^\pm$ lines in the Ising model have the following properties. (i) They separate three states, positive ferromagnetic-like, paramagnetic and negative ferromagnetic-like states (see Fig.~\ref{fig:schematic}B). 
%The two lines do not merge to a common line even in the vicinity of the critical point -- they coincide  precisely at the critical point.    (ii) Both lines emanate from the critical point, 
(ii)  The two lines coincide  only at the critical point. They do not merge to a common line even in the vicinity of the critical point.
%Both lines emanate from the critical point, and do not merge to a common line even in the vicinity of the critical point -- they coincide  precisely at the critical point.
(iii) The two lines follow the critical scalings Eqs.~(\ref{eq:scalingH}) and (\ref{eq:scalingm}).
%Comparing the crossover lines (1-6) in supercritical fluids and $L^{\pm}$ in the Ising model, one finds that the universality of critical phenomena is violated.
None of the above-mentioned supercritical crossover lines (I-{VII}) simultaneously satisfy all these three properties  (see SI Table~S1 for a summary).
Definitions (I-V) give one single crossover line; definition (VI) gives two percolation lines, but they do not emanate from the critical point.
The (II) Fisher-Widom line, (III) Frenkel line, and (VI) percolation lines do not pass through the critical point, and thus are without any critical scaling. The (I) Widom line, (IV) Nishikawa line, (V) symmetry line, and {(VII) pseudo-boiling boundaries $T^{\pm}(P)$}, although emanating from the critical point, do not satisfy the critical scalings Eqs.~(\ref{eq:scalingH}) and (\ref{eq:scalingm}). In short, even though it is well established that liquid-gas and ferromagnetic-paramagnetic critical points belong to the same 3D Ising universality class, their supercritical physics is not yet unified within the framework of critical universality.
%Among the six proposals, only (6) percolation lines have two lines  (see Table~\ref{table:lines} for a summary).

%The Fisher-Widom line and the Frenkel line do not pass through the critical point, and thus do not have any scaling laws.  The Widom line, although  emanating from the critical point, does not follow the scaling $|H| \sim (T - T_{\rm c})^{\beta+\gamma}$. Even more problematically, the number of crossover lines in the supercritical region do not match in two systems:  as revealed by their critical scalings, the two lines $L^{\pm}$ in the Ising model do not merge to a common line even in the vicinity of the critical point -- they coincide only exactly at the critical point.   

\section*{Methods and Results}
To resolve the inconsistency, we propose a new definition of the crossover lines in supercritical fluids (see Fig.~\ref{fig:schematic}C). The method  is explained in Fig.~\ref{fig:method}, taking argon as an example.
The general procedure is as follows. 
(i) The experimental data of 
 supercritical 
 %(for $\rho \leq \rho_{\rm c}$)
 isothermal equations of states (EOSs), $\hat{P}(\hat{\rho}; \hat{T})$, are collected from the National Institute of Standards and Technology (NIST) database~\cite{NIST} and plotted in Fig.~\ref{fig:method}(A), where the state variables are rescaled by their critical values,  $\hat{T} = T/T_{\rm c}$, $\hat{P} = P/P_{\rm c}$ and $\hat{\rho} = \rho/\rho_{\rm c}$.  (ii)  The critical isochore $\hat{P}(\hat{\rho} = 1; \hat{T})$  is subtracted from the original EOSs, giving the shifted EOSs (Fig.~\ref{fig:method}B), $\delta \hat{P}(\hat{\rho}; \hat{T}) = \hat{P}(\hat{\rho}; \hat{T}) - \hat{P}(1; \hat{T})$. 
    Here the critical isochore is  analogous to the $H=0$ line in the Ising model; in particular, 
    along the 
   critical isochore,  $\kappa_T$  exhibits the standard critical scaling,  $\kappa_T \sim (T-T_{\rm c})^{-\gamma}$, for $T>T_{\rm c}$.
   Thus the critical isochore above the critical point can be considered as an 
   extension of the coexistence line.
   Due to this symmetry, we take the  critical isochore as the reference line to obtain shifted EOSs.
   % \beta_T \sim \Delta \hat{T}^{-\gamma}$, where $\Delta \hat{T} = T/T_{\rm c} -1$.
%Next we work on the shifted EOSs.  
Similar to  the isothermal EOSs $H(m;T)$ of the Ising model, which all intersect at a point $( m=0, H=0 )$,
%{\color{red} (see SI XXX),}\YJ{this figure is missing.} 
the shifted EOSs $\delta \hat{P}(\hat{\rho}; \hat{T})$ intersect at a common point  $( \hat{\rho}=0, \delta \hat{P}=0 )$.
The susceptibility $\kappa_T$  
%of the order parameter $\hat{\rho}$ in response to  $\delta \hat{P}$, which 
is the inverse slope of the shifted EOS for constant $\hat{T}$.
The estimated $\kappa_T$ is consistent with the density fluctuation data measured in scattering experiments (see SI Fig.~S18 for the case of carbon dioxide).
%the critical point. 
 %Analogous to the Ising model, we consider $\delta \hat{P}$ as the external field, and define a susceptibility $\kappa_T \equiv \left( \frac{\partial \hat{\rho}}{\partial (\delta \hat{P})} \right)_{\hat{T}}   = \left( \frac{\partial \hat{\rho}}{\partial  \hat{P}} \right)_{\hat{T}}  =   \frac{P_{\rm c}}{\rho_{\rm c}}  \left(\frac{\partial \rho}{\partial  P} \right)_T$ of the order parameter $\hat{\rho}$ in response to  $\delta \hat{P}$, which is the inverse slope of the shifted EOSs (to avoid confusion, in this paper we call  $\kappa_T$  susceptibility and $\beta_T$  compressibility).  
   (iii)
   For a fixed $\delta \hat{P}$, the compressibility $\kappa_T(\hat{T}; \delta \hat{P})$ peaks at $\hat{T}_{\rm max}^{+}(\delta \hat{P})$ for $\delta \hat{P}>0$ (see Fig.~\ref{fig:method}C) and 
   $\hat{T}_{\rm max}^{-}(\delta \hat{P})$ for $\delta \hat{P}<0$  (see Fig.~\ref{fig:method}D).
   Connecting the points $\hat{T}_{\rm max}^{\pm}(\delta \hat{P})$ together gives the two lines $L^{\pm}$ (see Fig.~\ref{fig:method}B).
   Figure~\ref{fig:method}E shows $L^{\pm}$ lines
    in the  $\Delta\hat{P}- \Delta \hat{T}$ phase diagram, where $\Delta\hat{P} = P/P_{\rm c} -1$ and $\Delta\hat{T} = T/T_{\rm c} -1$.
   These two lines separate liquid-like, liquid-gas-indistinguishable and gas-like phases, which are counterparts of positive ferromagnetic-like,  paramagnetic and  negative ferromagnetic-like phases in the Ising model. The described procedure  applies generally to the van der Waals EOS, a standard theoretical model for gases and liquids (see SI Sec.~S3), and to the experimental EOSs of other substances (e.g., water, see SI Sec.~S4). 
 % \YJ{Is there an existing name for $\kappa_T$?}

  It is clear that $L^{\pm}$ lines are defined according to {\it thermodynamics} rather than dynamics. 
   The criterion of $L^{\pm}$ is similar to that of the Widom line; the key difference is that the maxima of the response function $\kappa_T$ are evaluated along  paths parallel  to the critical isochore (see Fig.~\ref{fig:method}E), instead of along constant-pressure or constant-temperature paths. Each parallel path that is $\delta \hat{P}$ away from  the critical isochore is tangential to a constant-$\kappa_T$ contour line, at the point of tangency $\hat{T}_{\rm max}^{+}(\delta \hat{P})$ or $\hat{T}_{\rm max}^{-}(\delta \hat{P})$. Note that for better visualization, in Fig.~\ref{fig:method}E we plot the color map and contour lines of constant $1/\kappa_T$, where a maximum becomes a minimum without changing its location.

\begin{figure*}[!htbp]
  \centering
  \includegraphics[width=\linewidth]{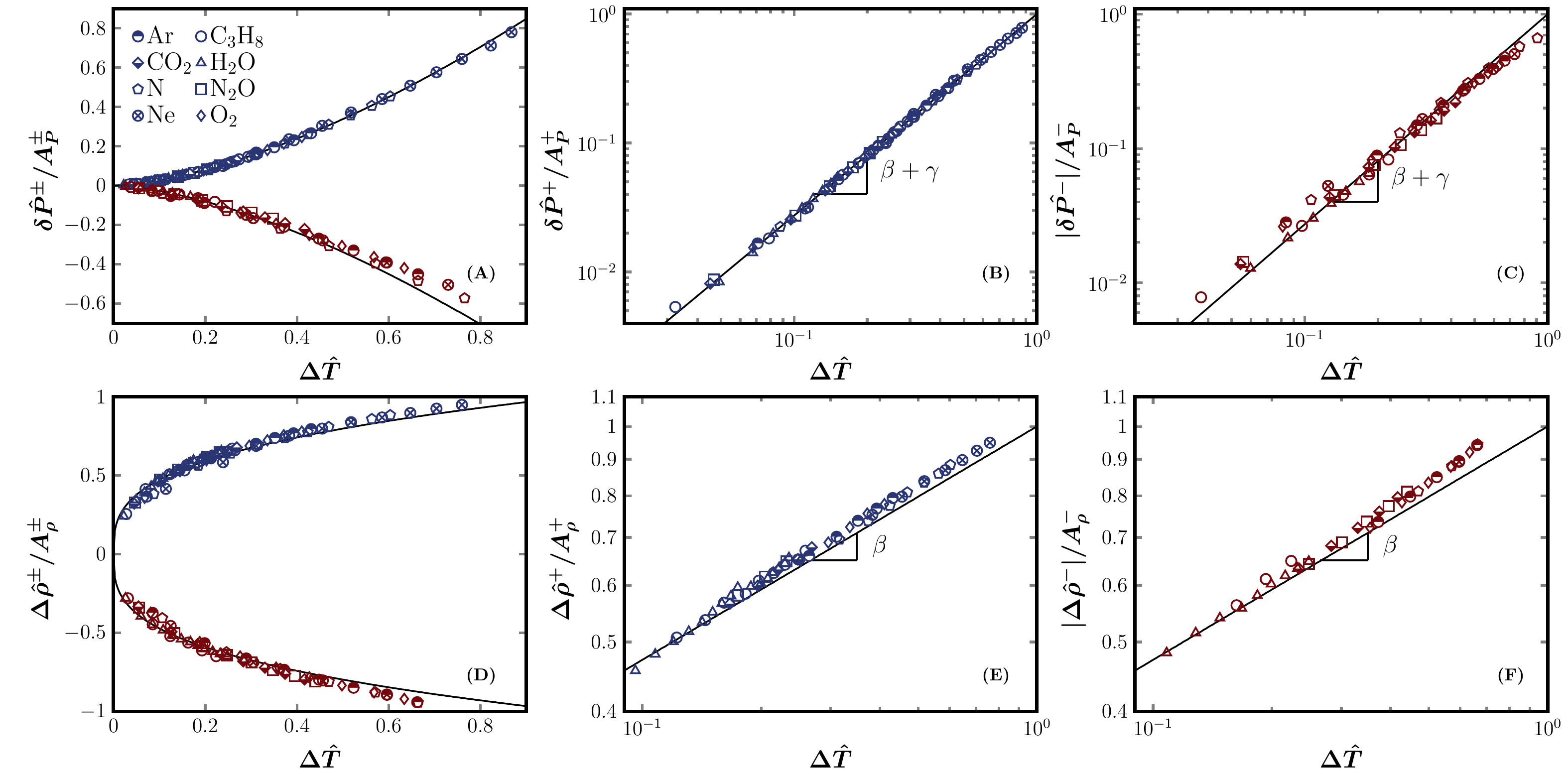}
  \caption{{\bf Scalings of supercritical crossover lines  $L^{\pm}$ near the critical point in eight substances.}
  %substrates.}
  (A)  $\delta \hat{P}^{\pm}/A_P^{\pm}$ as a function of $\Delta \hat{T}$.
  %We plot $\delta \hat{P}^{\pm}$ along $L^{\pm}$ lines, rescaled by $A^{\pm}$, as functions of $\Delta \hat{T}$. 
  The data points are fitted to $|\delta \hat{P}^{\pm}| = A_{P}^{\pm} \Delta \hat{T} ^{\beta + \gamma}$ (solid lines), where $A_{P}^{\pm}$ are non-universal fitting parameters (see SI Table S2 for values of $A_{P}^{\pm}$), and the exponents are fixed by the 3D Ising universality values, $\beta = 0.3265$ and $\gamma = 1.237$.
  %\YJ{In A and D, change y-axis label to $\delta \hat{P}^{\pm}/A_P^{\pm}$}
 (B) and (C) are log-log plots for $L^+$ and $L^-$ respectively. 
 % \YJ{Notations to be avoid a confusion  with $C_P$. For Ising model, we can use $\delta \hat{H}^{\pm} = A_{H}^{\pm} (T-T_{\rm c})^{\beta + \gamma}$, etc. Draw fitting lines in A and D. Change axis labels to be consistent. Make the change everywhere.}
   (D-F) Similar plots for the scaling  $|\Delta \hat{\rho}^{\pm}| = A_{\rho}^{\pm} \Delta \hat{T}^{\beta}$.
   %\YJ{In B and E, remove the absolute sign, use $\delta \hat{P}^+/A_P^+$, In C, F, use $|\delta \hat{P}^-|/A_P^-$, make sure it is $A_P^+$ not $A_{P^+}$ everywhere}
   %\YJ{add fitting lines in (D)}
  %Collapse of $L^{\pm}$ for some common substrates (Ar, $\mathrm{CO}_2$, N, Ne, $\mathrm{H}_2\mathrm{O}$, etc.) in (A)  $\delta \hat{P}-\Delta \hat{T}$ phase diagram and (D) $\Delta \hat{\rho}-\Delta \hat{T}$ phase diagram, by dividing $|\delta \hat{P}|$, $|\Delta \hat{\rho}|$ by constant coefficients $c_P$, $c_\rho$, which are the coefficients of Eq.~(\ref{eq:scaling}) and Eq.~(\ref{eq:scaling2}). The values of $c_P$ and $c_\rho$ are presented in SI. \YJ{We have to give the value of C. make a table in SI to present the values?} (B) and (C) are double logarithmic forms of $L^+$ and $L^-$ in $\delta \hat{P}-\Delta \hat{T}$ phase diagram, (E) and (F) are double logarithmic forms in $\Delta \hat{\rho}-\Delta \hat{T}$ phase diagram, and black lines are fitted lines of these data, according to Eq.~(\ref{eq:scaling}) and Eq.~(\ref{eq:scaling2}).
  %(add other common substrates, e.g., water, CO2, etc.)
  }
  \label{fig:scaling}
\end{figure*}

   Using a linear scaling theory (see SI Sec.~S5)~\cite{luo2014behavior}, we establish theoretically the  scalings of  $L^{\pm}$ lines near the liquid-gas critical point,
   \beq
   |\delta \hat{P}^{\pm}|  \sim (T - T_{\rm c})^{\beta + \gamma},
   \label{eq:scaling}
   \eeq
   and
      \beq
   |\Delta \hat{\rho}^{\pm}|  \sim (T - T_{\rm c})^{\beta},
   \label{eq:scaling2}
   \eeq
   where $\delta \hat{P}^{\pm}$ and $\Delta \hat{\rho}^{\pm}$ are values on the  $L^{\pm}$ lines ($\Delta \hat{\rho} = \rho/\rho_{\rm c} - 1$). To verify Eqs.~(\ref{eq:scaling}) and~(\ref{eq:scaling2}), we collect the experimental data of eight common substances, including water ($\rm{H_2O}$), carbon dioxide ($\rm{CO_2}$), argon ($\rm{Ar}$), nitrogen ($\rm{N}$), neon ($\rm{Ne}$), propane ($\rm{C_3H_8}$), dinitrogen monoxide ($\rm{N_2O}$) and oxygen ($\rm{O_2}$) from the NIST database, and evaluate their $L^{\pm}$ lines using the same approach as demonstrated above for argon (Fig.~\ref{fig:method}). Figure~\ref{fig:scaling} shows that both scalings are confirmed by the data of all eight substances examined,  with the  3D Ising universality class exponents.
Furthermore, the scalings Eqs.~(\ref{eq:scaling}) and~(\ref{eq:scaling2}) are robust, independent of which thermodynamic response function is chosen to determine the maxima (see SI Sec.~S5 and~S7).
   Equations~(\ref{eq:scalingH}-\ref{eq:scaling2}) recover the critical universality: the two supercritical crossover lines $L^{\pm}$ 
   obey scaling laws with the same exponents in both liquid-gas and Ising systems.

\begin{figure*}[!htbp]
  \centering
  \includegraphics[width=\linewidth]{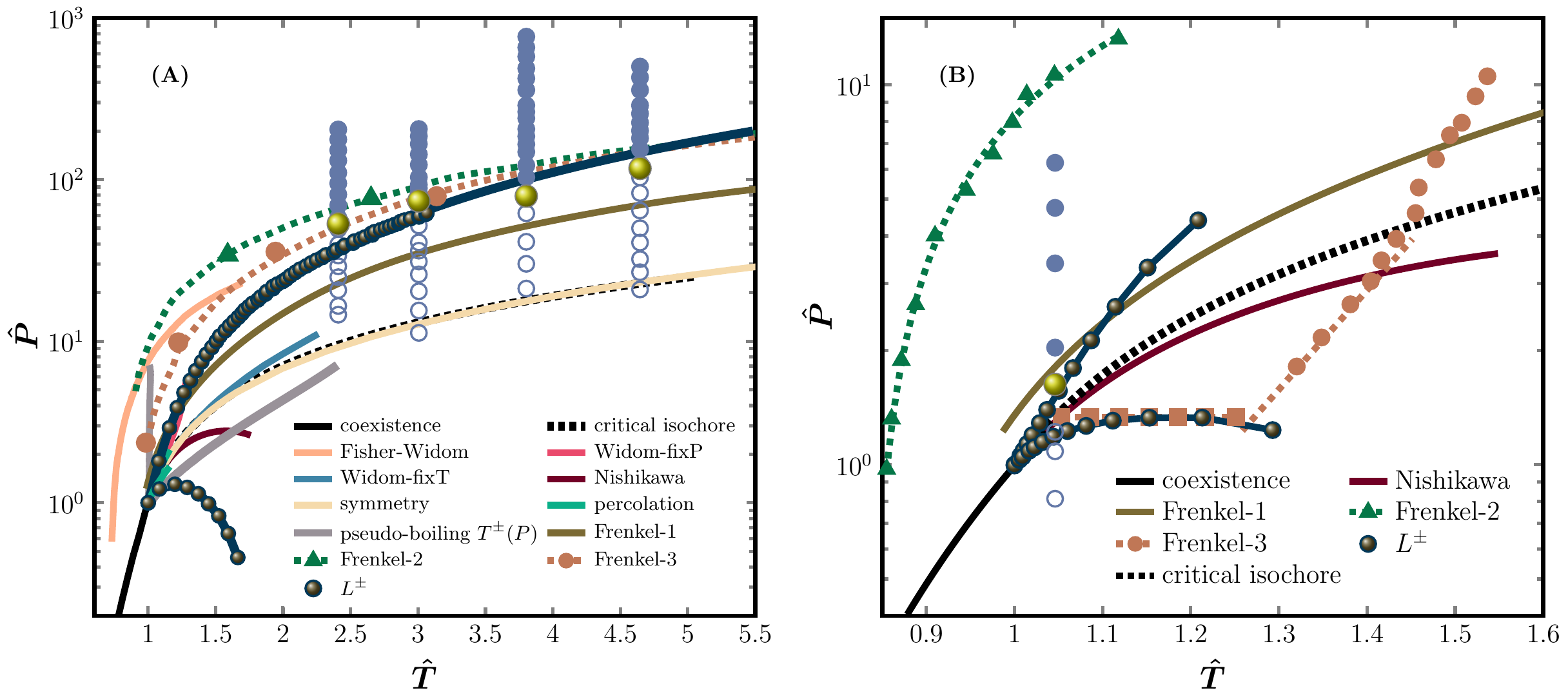}
  \caption{%{\bf Validations of $L^+$.} 
  {\bf Comparison between supercritical crossover lines and experimental (simulation) data. }
  (A) $\hat{P}$-$\hat{T}$ diagram of supercritical argon. 
  %Validation of $L^+$ with experimental data of Argon. 
  %The critical isochore (dashed black line) is extended from the coexistence line (solid black line)~\cite{simeoni2010widom}.
  The data points of $L^{\pm}$ lines are obtained in Fig.~\ref{fig:method}, and fitted to the scaling form Eq.~(\ref{eq:scaling}) (solid lines, see SI Sec.~S7 for a discussion on the fitting).
  %dark blue line with gray balls on it is $L^{\pm}$, which is the only one that coincide with the experimental data.
  Filled and open blue circles are  data points collected from ~\cite{simeoni2010widom, gorelli2013dynamics} at four different $\hat{T}$: the filled  circles are with 
positive increasing sound dispersion,  and the open circles are with weak constant sound dispersion; they are separated by crossover points indicated by yellow balls (see SI Sec.~S8).
  %which can be seen as a feature of liquid-like regime, and hollow dots belongs to another regime. Yellow balls are their boundaries.
  %\YJ{reorder Frankel 1, 2, 3 in B, change Frankle to Frankel-1, also Fig. S15}
  The Fisher-Widom line ~\cite{gallo2016water}, 
  %the Frenkle line for a Lennard Jones fluid~\cite{brazhkin2012two} (Frenkel-1), 
  Frenkel-1 line~\cite{brazhkin2012two}, Frenkel-2 line~\cite{brazhkin2013liquid}, Frenkel-3 line~\cite{brazhkin2012two},
  symmetry line~\cite{ploetz2019gas} and percolation lines~\cite{woodcock2013observations} are copied from the cited publications correspondingly.
  The Nishikawa line is evaluated in SI Sec.~S10 according to its definition. 
  The Widom lines are determined based on the NIST EOSs, for fixed $P$ and  fixed $T$.
  {The pseudo-boiling boundaries $T^{\pm}(P)$ are determined following the definition given in~\cite{banuti2015crossing}.}
  An enlarged view near the critical point is provided in Fig.~S15.
%The light orange, claybank, purple, buff and turquoise lines represent the Fisher-Widom line ~\cite{gallo2016water}, Frenkle line~\cite{brazhkin2012two}, Nishikawa line, Symmetry line~\cite{ploetz2019gas} and percolation line~\cite{woodcock2013observations}. The pink and blue lines are widom lines calculated at fixed P and fixed T respectively. \YJ{add fixed-P Widom line.} 
(B) $\hat{P}$-$\hat{T}$ diagram of supercritical carbon dioxide. 
%Another validation of $L^+$ through the small-angle neutron scattering data of $\mathrm{CO}_2$, 
 Filled and open blue circles are  data points collected from ~\cite{pipich2018densification}: the filled/open circles are with/without enhanced density fluctuations compared to the NIST data, separated by the crossover point (yellow ball).
%\blue{In both panels, the Frenkel line depends on the criterion used:  Frenkel-1 depends on $\tau \approx \tau_0$. Frenkel-2( from \cite{yang2015frenkel}) is decided by the disappearance of solid-like vibration motion, which can be extracted from the qualitative change in the time behavior of velocity autocorrelation from oscillation to monotonically decreasing, or from the self-intermediate scattering function $F_s(q,t)$, where the maxima of $F_s(q,t)$ at short time will disappear~\cite{brazhkin2012two}. Frenkel-3 (from \cite{fomin2015thermodynamic,  pipich2018densification}) is derived from isochoric heat capacity when it is equal to $3.5k_B$ for $\rm CO_2$. While for monatomic molecules like $\rm A_r$, the criterion is $C_v=2k_B$.}
  The Frenkel-1 line~\cite{brazhkin2012two}, Frenkel-2 line~\cite{yang2015frenkel}, Frenkel-3 line (defined at $C_V=3.5 k_{\rm B}$ for carbon dioxide)~\cite{fomin2015thermodynamic,  pipich2018densification},
  and Nishikawa line~\cite{imre2015anomalous}
  are copied from the cited publications correspondingly.
 %The Nishikawa line is from~\cite{imre2015anomalous}.
 %\YJ{In the legend, replace the line by points for $L^\pm$, add a triangle to dashed line for Frenkel-2, a circle to the dashed line for Frenkel-3}
%which are also represented by solid dots and hollow dots for two different regime~\cite{pipich2018densification}, and yellow ball is the boundary. \blue{We add critical isochore and $L^{\pm}$ to FIG.1 in ~\cite{pipich2018densification}, where the black dashed line is critical isochore, and the dark blue dotted line is $L^{\pm}$. The green Frenkle line is calculated from the velocity autocorrelation function, the claybank Frenkle line is on the basis of a Lennard Jones fluid, which is the same one in (A), and the orange one with squares and circles is derived from isochoric heat capacity. Actually, the one which is named Widom line in ~\cite{pipich2018densification} is Nishikawa line. All these four lines can be find in ~\cite{pipich2018densification}.}
  }
  \label{fig:experiment_L+}
\end{figure*}

Next we compare $L^{\pm}$ lines with supercritical crossovers reported previously in two sets of independent experiments (simulations)~\cite{simeoni2010widom, gorelli2013dynamics, pipich2018densification}.  
%Next we verify the definition of $L^{\pm}$ by independent experimental and simulation data. For $L^+$, we compare our result with the data 
In the first set, {\it dynamical} supercritical crossovers are estimated based on the sound dispersion data obtained for supercritical argon, by inelastic X-ray scattering experiments and molecular dynamics simulations~\cite{simeoni2010widom, gorelli2013dynamics}.  From these data,  sharp crossovers are determined between a regime of  positive sound dispersion ratio that increases with $\rho$ at constant $T$, and a regime of weak, $\rho$-independent sound dispersion (see SI Sec.~S8).
The increasing positive dispersion in the first regime reflects typical viscoelastic behavior of liquid-like states, implying the existence of two relaxation mechanisms --  structural relaxations related to the so-called $\alpha$-processes and  microscopic relaxations related  to nearest-neighbour interactions~\cite{simeoni2010widom, gorelli2006liquidlike, bencivenga2006adiabatic}.
In the second regime, the weak constant dispersion implies that the structural relaxations disappear and only the microscopic relaxations remain, representing non-liquid-like  (gas-like or indistinguishable) states~\cite{ruocco2000relaxation, scopigno2002evidence}.
%former behavior is originated from the  structural relaxation ($\alpha$-processes) and  viscoelastic characteristic  of liquid-like states~\cite{simeoni2010widom, gorelli2006liquidlike, bencivenga2006adiabatic};the latter can be attributed  to microscopic relaxations related to nearest-neighbour interactions, which should present universally in different thermodynamic states~\cite{ruocco2000relaxation, scopigno2002evidence}. 
In Fig.~\ref{fig:experiment_L+}(A), the dynamical crossovers from sound-dispersion data are compared with $L^{\pm}$ lines and previously proposed crossover lines (I-VI). Good agreement is found between the experimental dynamical  crossovers and the $L^+$ line.
Interestingly, the $L^+$ line is close to the Frenkel line determined by various criteria.
In Fig.~\ref{fig:experiment_L+}, the
Frenkel-1 line is defined by the equality $\tau_0 = \tau$ between the vibration time $\tau_0$ and the liquid relaxation time $\tau$ for the Lennard-Jones model~\cite{brazhkin2012two}, the Frenkel-2 line corresponds to the disappearance of oscillations and minima of the velocity autocorrelation function~\cite{brazhkin2013liquid}, and the Frenkel-3 line is derived based on isochoric heat capacity  
$C_V$ (for monatomic systems such as argon, the criterion is $C_V= 2k_{\rm B}$, where $k_{\rm B}$ is the Boltzmann constant)~\cite{brazhkin2012two}.
We emphasize that the Frenkel line has a dynamic nature, and it cannot satisfy the scalings Eq.~(\ref{eq:scaling}) and~(\ref{eq:scaling2})  in the vicinity of the critical point since it 
does not end at the critical point (see Fig.~S15).
In future studies, it is of interest to understand if there is a deep connection between $L^+$ and Frenkel lines, which are  based respectively  on thermodynamic and dynamic criteria.

%\blue{In addition, the Frenkel line-1 for a Lennard Jones fluid is very close to $L^+$ and also quite good in seperating positive dispersion region. The line is decided by the time when the relaxation time $\tau$ of system is close to the vibration time $\tau_0$ of an atom. It behaves very closely to the $L^+$ at high temperature and pressure, and the most significant difference between them is that the Frenkle line does not cross the critical point (see SI FIG.~S15). Therefore, we need more precise experimental data, especially those near the critical point, to tell the difference when comparing to the experiments.}

%We emphasize the {\it dynamic} nature of the crossovers determined from sound dispersion data, because they reflects the change of transport properties and relaxation mechanisms. Their agreement with $L^+$ line thus represents a remarkable coincidence between dynamic and thermodynamic supercritical crossovers.

In the second experiment, a supercritical crossover is estimated based on
%Another experimental support is found by comparing $L^+$ line to 
the small-angle neutron scattering data of carbon dioxide~\cite{pipich2018densification}. The data show that, above a crossover pressure, { $\hat{P}_{\rm cro} = 1.63$ ($P_{\rm cro} =120$ bar)
%at {\color{red}  $P/P_{\rm c} = 0.81-6.23$ 
at a fixed $\hat{T} = 1.046$ ($T=45 ^{\circ}$ C)},
the density and density fluctuation are enhanced   
%of supercritical carbon dioxide $\mathrm{CO}_2$ 
compared with the standard NIST data for a homogeneous state.
It is expected that the formation of liquid droplets causes such enhancement; correspondingly $\hat{P}_{\rm cro}$ indicates a crossover to liquid-like behavior.  
The location of this crossover point  is consistent with $L^+$ line, as shown in Fig.~\ref{fig:experiment_L+}(B). 

%In Ref.~\cite{ploetz2019gas}, besides a symmetry line dividing liquid-like and gas-like behavior in supercritical fluids, the authors also define two boundary lines of the transition region between the two kinds of behavior.At the boundary line, the quadruplet  particle number fluctuation density reaches a maximum along a particular isotherm. We find that these two boundary lines follow the same critical scaling Eq.~(\ref{eq:scaling}) as $L^{\pm}$, but deviate from $L^{\pm}$ at larger pressures (see Fig.~\ref{fig:experiment_L+}A). Conceptually, we emphasize that the region between $L^+$ and $L^-$ should not be interpreted as a "transition region" that broadens a single line (e.g., the symmetry line) dividing liquid-like and gas-like behavior; it should be instead understood as a single phase where liquids and gases are indistinguishable, similar to the paramagnetic phase in the Ising model. The sound dispersion data~\cite{simeoni2010widom, gorelli2013dynamics} favor the latter: the dispersion is a constant in this region, without any noticeable  change across the symmetry line (see SI). 

\section*{Discussion}
The above two examples demonstrate remarkable coincidences between the $L^+$ line and previous experimental results. The prediction of $L^-$ line remains to be examined in future experiments. In SI Sec.~S11, we show that the $L^-$ line can be validated independently by the behavior of EOS for the compressibility factor $Z$ above the critical temperature: the EOS of gas states  can be well described by a universal form $Z_{\rm gas}(\rho, T)$ derived from the  van der Waals equation in the dilute gas limit, 
and the deviation points from 
$Z_{\rm gas}(\rho, T)$
coincide with the $L^-$ line.
Based on the above results,  
the physical meaning of two supercritical crossover lines 
can be interpreted. 
The $L^+$ and $L^-$ lines represent the loci in the phase diagram where the system starts to deviate from  liquid and gas behavior respectively, and the indistinguishable states between the two lines behave neither like a standard liquid nor  a standard gas.
This study focuses on the thermodynamic consequences of the $L^+$ and $L^-$ lines.
From the dynamic viewpoint, the deviation from the liquid dynamics corresponds to the Frenkel line, where the vibration motion of molecules disappears. As shown in Fig.~\ref{fig:experiment_L+}A, the $L^+$ and Frenkel lines are indeed close to each other.
The deviation from the gas dynamics has not been considered previously; it will be interesting to investigate how the dynamics change around the $L^-$ line in the future.

%However, we expect that the differences  between {gas-like} and indistinguishable states might be subtle to detect directly in experiments.  

Finally, it should be straightforward to generalize the present analysis to other phase transitions, including liquid-liquid phase transitions~\cite{xu2005relation}, quantum phase transitions~\cite{jimenez2021quantum}, non-equilibrium phase transitions such as liquid-glass transitions~\cite{berthier2015evidence}, and phase transitions in dusty plasmas~\cite{huang2023revealing}. {In particular, recent studies~\cite{stapmanns2018thermal, weber2022quantum, jimenez2021quantum, wang2023plaquette} have revealed  the analogy between the critical point of water and that 
of certain quantum magnetic systems, where two supercritical crossover lines can be determined according to the behavior of specific heat. It is expected that the $L^{\pm}$ lines in such  quantum magnetic systems can be defined following the method provided here, which should obey the same universal scalings as Eqs.~(\ref{eq:scaling}) and~(\ref{eq:scaling2}), independent of the slope of the coexistence line (see SI Sec.~S5) and of which thermodynamic response function is chosen to determine $L^{\pm}$ (see SI Sec.~S5 and~S7B).}

\begin{acknowledgments}
We thank Limei Xu, Matteo Baggioli, Yan Feng, Xinzheng Li, Haijun Zhou, Chengran Du, Jie Zhang, Rui Shi and Chong Zha for discussions. We acknowledge financial support from NSFC (Grants 11935002,
11974361, 12161141007 and 12047503), from 
Chinese Academy of Sciences (Grants ZDBS-LY-7017 and KGFZD-145-22-13), 
and from Wenzhou Institute (Grant WIUCASICTP2022). In this work access was granted to the High-Performance Computing Cluster of Institute of Theoretical Physics - the Chinese Academy of Sciences.

\end{acknowledgments}

\bibliography{widom}
\onecolumngrid
%\appendix

\clearpage
%\onecolumn

\centerline{\bf \Large Supplementary Information}
\tableofcontents
\setcounter{figure}{0}
\setcounter{equation}{0}
\setcounter{table}{0}
\renewcommand\thefigure{S\arabic{figure}}
\renewcommand\theequation{S\arabic{equation}}
\renewcommand\thesection{S\arabic{section}}
\renewcommand\thetable{S\arabic{table}}

\section{Ising model}
The Hamiltonian of the Ising model is~\cite{BRUSH1967History}, 
%classical nearest-neighbor Ising Hamiltonian is~\cite{BRUSH1967History} 
\beq
\mathcal{H}=-J\displaystyle\sum_{\left<i,j\right>}s_is_j-H\sum_{i=1}^{N}s_i,
\label{eq:Ising Hamiltonian}
\eeq
where $J$ is a coupling constant, $s_i = \pm 1$ the value of spin on the $i$-th site, $N$ the total number of spins, $H$  the  external magnetic field, and $\left<i,j\right>$ stands for nearest-neighbor spin pairs. 
The model is defined on a square lattice  in $d$ dimensions. 
The mean magnetization is $m = \frac{1}{N} \left \langle \sum_{i=1}^N s_i \right \rangle $, where $\left \langle \ldots \right \rangle$ is the ensemble average,  and the susceptibility is $\chi = \left( \frac{\partial m}{\partial H} \right)_{T}$. 

According to the definition, on the supercritical ($T > T_{\rm c}$) crossover lines $L^{\pm}$, $\chi$ is maximum, i.e., 
\beq
\left( \frac{\partial \chi}{\partial T} \right)_{H} = 0,
\eeq
 where  $T_{\rm c}$ is the critical temperature. The purpose of this section is to show  that on $L^{\pm}$ the order parameter $m^{\pm}$ and the field $H^{\pm}$ satisfy two scalings,  Eqs.~(1) and~(2), near the critical point.
% \beq
%\left | m^{\pm}  \right | \sim \left (T  - T_{\rm c} \right )^{\beta},
%\label{eq:scaling_m}
% \eeq
% where $\beta $ is a standard critical exponent  in the scaling $m \sim (T_{\rm c}-T)^\beta$,
% and
%\beq
%\left | H^{\pm}\right |  \sim \left (T  - T_{\rm c}\right)^{\beta + %\gamma},
%\label{eq:scaling_h}
%\eeq
%where $\gamma $ is another standard critical exponent  in the scaling $
%\chi\sim \left| T - T_{\rm c} \right)^\gamma$.  
We first derive these two scalings analytically from a mean-field calculation, and then demonstrate that they also hold in two and three dimensions ($d=2$ and 3) using Monte Carlo simulations.

\subsection{Mean-field theory}

\begin{figure}[!htbp]
  \centering
  \includegraphics[width=\linewidth]{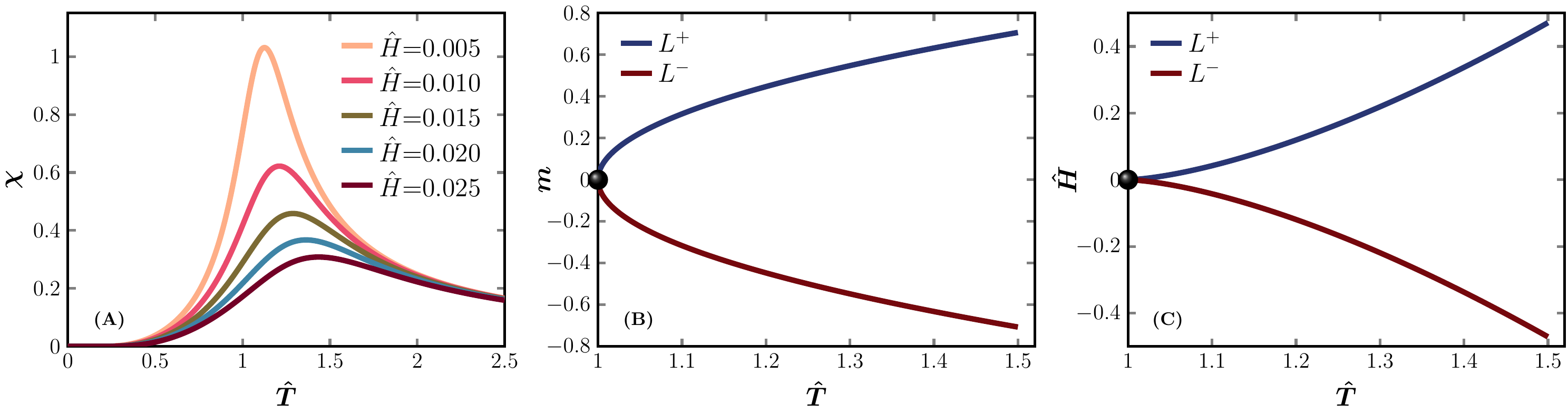}
  \caption{{\bf Supercritical crossover lines $L^{\pm}$ in the mean-field Ising model.}
  (A) Magnetic susceptibility $\chi$ as a function of the reduced temperature $\hat{T}$, for a few different reduced magnetic fields $\hat{H}$. The peaks determine the supercritical crossover lines $L^{\pm}$. 
  Along $L^{\pm}$,  (B) $m^{\pm}$  changes with $\hat{T}$ following Eq.~(\ref{eq:MF_scaling_m}), and (C) $\hat{H}^{\pm}$ changes with $\hat{T}$  following Eqs.~(\ref{eq:MF_scaling_h}).
  %\YJ{$\hat{H}$, $L^+$}
  }
  \label{fig:Ising}
\end{figure}

With the mean-field approximation, $m$ and $\chi$ are easy to calculate~\cite{baxter2016exactly}:
%\YJ{add a citation, maybe a standard book}:
%, in which case
\beq
m = \tanh{\left[\beta (H + qJm) \right]},
\label{eq:magnetization}
\eeq
and 
\beq
\chi = \frac{\beta} {\cosh^2{\left[\beta (H + qJm)\right] - \beta Jq}},
\label{eq:suceptibility}
\eeq
 where $q = 2d$ is the coordination number (i.e., the number of nearest neighbors of each spin), and $\beta = 1/{k_{\rm B}T}$  the inverse temperature. We set the Boltzmann constant $k_{\rm B} = 1$.

In the following derivations, we switch to  the reduced parameters  $\hat{H} = H/T_{\rm c}$  and 
 $\hat{T}= T/T_{\rm c}$, where $T_{\rm c} = q J$.
 %where $T_{\rm c} = q J$ is the critical temperature. 
%This relation can be also derived by mean filed approximation easily. If we define $t=T/T_{\rm c}, b=h/T_{\rm c}$, and substitute $\beta=1/T, T_{\rm c}=qJ$ into 
With the reduced parameters, Eq.~(\ref{eq:magnetization}) becomes,
 \beq
 \begin{aligned}
m &= \tanh{\left[\frac{ (\hat{H} + m)}{\hat{T}} \right]}\\
  &\approx \frac{\hat{H}+m}{\hat{T}}-\frac{1}{3} \left(\frac{\hat{H}+m}{\hat{T}}\right)^3,  
  \label{eq:magnetization_3rd}
 \end{aligned}
\eeq
where we have expanded it up to the third order, assuming that $\hat{H}$ and $m$ are small near the critical point. Defining $\hat{\chi} = T_{\rm c} \chi$,  from Eq.~(\ref{eq:magnetization_3rd}) we obtain, 
 \beq
 \begin{aligned}
\hat{\chi}  =  \left ( \frac{\partial{m}}{\partial{\hat{H}}} \right )_{\hat{T}} = \frac{1-(\hat{H}+m)^2/\hat{T}^2}{\hat{T}-1+(\hat{H}+m)^2/\hat{T}^2}.
  \label{eq:suceptibility_3rd}
 \end{aligned}
\eeq
Making a change of variable $A = (\hat{H}+m)/\hat{T}$, and taking the partial derivative of Eq.~(\ref{eq:suceptibility_3rd}) with respect to the temperature, we get
 \beq
 \begin{aligned}
\left( \frac{\partial \hat{\chi}}{\partial \hat{T}} \right)_{\hat{H}} = \frac{(3A^2-1)(\hat{T}-1+A^2)+2A^2(1-A^2)}{(\hat{T}-1+A^2)^3},
  \label{eq:suceptibility_derivative}
 \end{aligned}
\eeq
which should be $0$ on the $L^{\pm}$.
The solution of $\left( \frac{\partial \hat{\chi}}{ \partial \hat{T}} \right)_{\hat{H}}= 0$ is, $(A^{\pm})^2 = \hat{T} - 1 $, or,
\beq
\left | m^{\pm}  \right |= \left (\hat{T} - 1 \right )^{1/2},
\label{eq:MF_scaling_m}
\eeq
near the critical point, where the exponent is equal to the mean-field exponent $\beta = 1/2$
(note that $\left | \hat{H}^{\pm}\right |$ is higher order; see Eq.~\ref{eq:MF_scaling_h}). Equation~(\ref{eq:MF_scaling_m}) gives the scaling relationship between the order parameter and temperature on $L^{\pm}$ (see Fig.~\ref{fig:Ising}). 
%\YJ{add a panel in Fig.S1}
Using Eq.~(\ref{eq:magnetization_3rd}), 
%\YJ{Eq. (S6) or Eq. (S5)? A:magnetization}
we can convert Eq.~(\ref{eq:MF_scaling_m}) into a relationship between the external field $\hat{H}^{\pm}$ and $\hat{T}$ on  $L^{\pm}$,
\beq
\left | \hat{H}^{\pm}\right |  =  \frac{4}{3} \left (\hat{T} - 1 \right)^{3/2},
\label{eq:MF_scaling_h}
\eeq
where the exponent is equal to $\beta + \gamma  = 3/2$ with the mean-field exponents $\beta = 1/2$ and $\gamma = 1$. The mean-field scalings are plotted in Fig.~\ref{fig:Ising}.
%is another standard mean-field exponent appearing in the scaling $ \chi\sim \left| 1-\hat{T} \right)^\gamma$.

Equations~(\ref{eq:MF_scaling_m}) and (\ref{eq:MF_scaling_h}) are the mean-field form of scalings Eqs.~(1) and~(2).
In Sec.~\ref{sec:linear scaling theory}, we will derive these scalings  using a general linear scaling theory without the mean-field approximation.

%\YJ{Make three panels in Fig.S1. (A) $\chi$ vs. $\hat{T}$ for a few different $\hat{h}$, similar to the current B. (B) Eq. (S8). (C) Eq. (S9), similar to the current A, but using only lines and don't distinguish between 2D and 3D.}

%Equations~(\ref{eq:scaling_m}) and~(\ref{eq:scaling_h}) can be written as, then %$m = (t-1)^{1/2}$ since $b<<m$ and $t\approx1$ near the critical point. %Finally, from Eq.~(\ref{eq:magnetization_3rd}),
% \beq
% \begin{aligned}
%b &= (t-1)m + \frac{1}{3} m^2\\
%  &= \frac{4}{3} (t-1)^{\frac{3}{2}},
%  \label{eq:relation}
% \end{aligned}
%\eeq
%scaling relationship is thus proved.

% We numerically calculate the magnetic susceptibility of the mean field 2D and 3D Ising model under fixed external magnetic fields. Fig.~\ref{fig:Ising}(B) shows the susceptibility curves of the 2D model under different external magnetic fields. The peak value of the curve decreases as the field strength increases, and the corresponding temperature increases. All the peaks will be combined into a line in the $H-T$ phase diagram, corresponding to the upper half of the curve in Fig.~\ref{fig:Ising}(A). Similarly, we calculate that under negative external magnetic fields, and in 3D model. The curves in different dimensions will coincide with each other through the transformation of $H/T_{\rm c}$ and $T/T_{\rm c}$, as shown in Fig.~\ref{fig:Ising}(A). These are  two crossover lines $L^{\pm}$ of the Ising model, which originate from critical point, and follow the scaling relation $|H| \sim (T - T_{\rm c})^{\beta+\gamma}$.

\subsection{Monte Carlo simulations in two and three dimensions}
%\YJ{In each dimension, show two figures. First figure (similar to Fig.1S): (A) $\chi$ vs $\hat{T}$; (B) $m^*$ vs $\hat{T}$ linear-scales; (C) $\hat{h}^*$ vs $\hat{T}$ linear-scales. Second figure (similar to Fig. 4(B-C)): (A) $m^*$ vs $\hat{T}-1$ log-scales; (B) $\hat{h}^*$ vs $\hat{T}-1$ log-scales.}

\begin{figure}[!htbp]
  \centering
  \includegraphics[width=\linewidth]{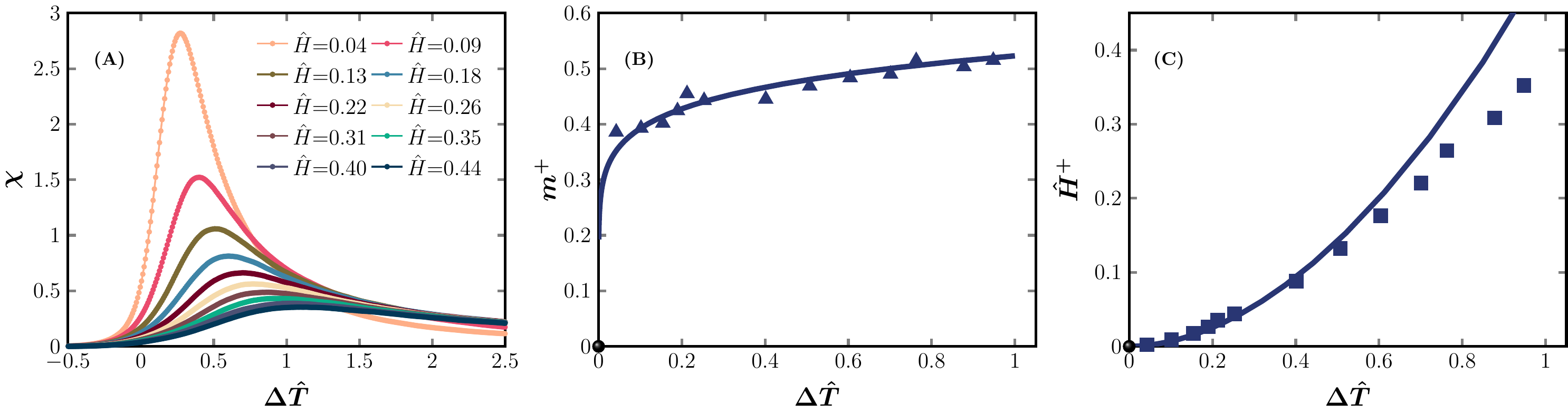}
  \caption{{\bf Supercritical crossover lines $L^{\pm}$ in the 2D Ising model.} Results are obtained from Monte Carlo simulations.
(A) Magnetic susceptibility $\chi$ as a function of the reduced temperature $\hat{T}$, for a few different reduced magnetic fields $\hat{H}$. The peaks determine the supercritical crossover line $L^{+}$ (the $L^{-}$ line is symmetric to the $L^{+}$ line and not shown). 
(B) The magnetization $m^{+}$ and (C) reduced field $\hat{H}^+$ along the $L^{+}$ line are plotted as functions of 
$\Delta \hat{T}$. 
%\YJ{unify $\hat{T}$, $\Delta \hat{T}$ in all plots.}
The lines represent fits, $m^+ = A_m \Delta \hat{T} ^\beta$ and $\hat{H}^+ = A_H \Delta \hat{T} ^{\beta+\gamma}$, where $A_m = 0.523$ and $A_H= 0.521$ are fitting parameters (see Fig.~\ref{fig:Ising_MC_2D_log} for corresponding log-log plots).}
%\YJ{convert b to $c_m$. Can you add one or two points closer to the critical point? In particular for m} }
%(A) Magnetic susceptibility $\chi$ as a function of temperature $\hat{T}$, for a few different magnetic fields $\hat{H}$. (B) $L^+$ of 2D Ising model in $m-\Delta{\hat{T}}$ phase diagram, and (C) in $\hat{H}-\Delta{\hat{T}}$ phase diagram.}
  \label{fig:Ising_MC_2D}
\end{figure}

\begin{figure}[!htbp]
  \centering
  \includegraphics[width=0.7\linewidth]{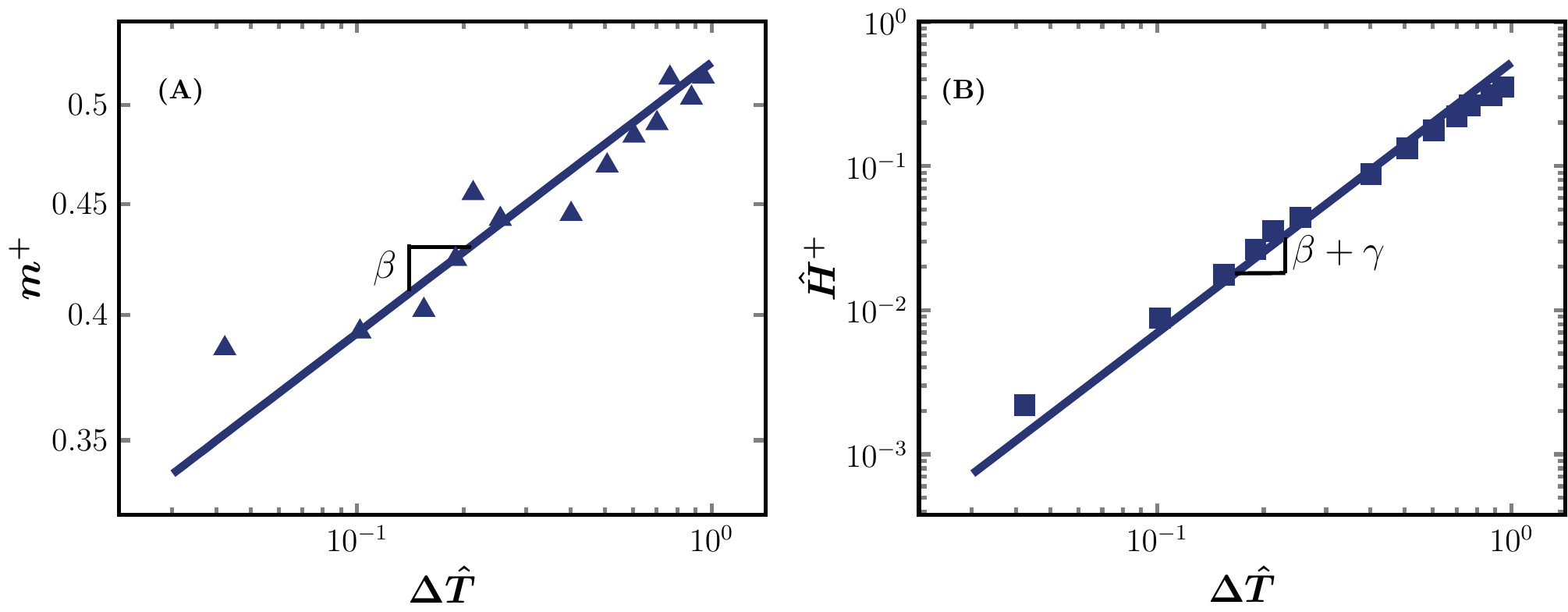}
  \caption{
 {\bf  Scaling laws  of the $L^+$ line in the 2D Ising model.} 
  (A) The magnetization $m^{+}$ and (B) reduced field $\hat{H}^+$  as functions of $\Delta \hat{T}$, in log-log scales. 
The lines represent fits, $m^+ = A_m \Delta \hat{T} ^\beta$ and $\hat{H}^+ = A_H \Delta \hat{T} ^{\beta+\gamma}$, where $A_m = 0.523$ and $A_H= 0.521$ are fitting parameters. 
%\YJ{Why are these numbers different from those in Fig. S2? Check carefully. Same for Figs. S3 and S4.}
%The points are fitting by f(x) = βx + b, where b = −0.55. (B) L+ of 2D Ising model in log-log scaled Hˆ − ∆Tˆ phase diagram. The points are fitting by f(x) = (β + γ)x + b, where b = −0.91. Solid lines are fitting curves.
  }
  \label{fig:Ising_MC_2D_log}
\end{figure}

\begin{figure}[!htbp]
  \centering
  \includegraphics[width=\linewidth]{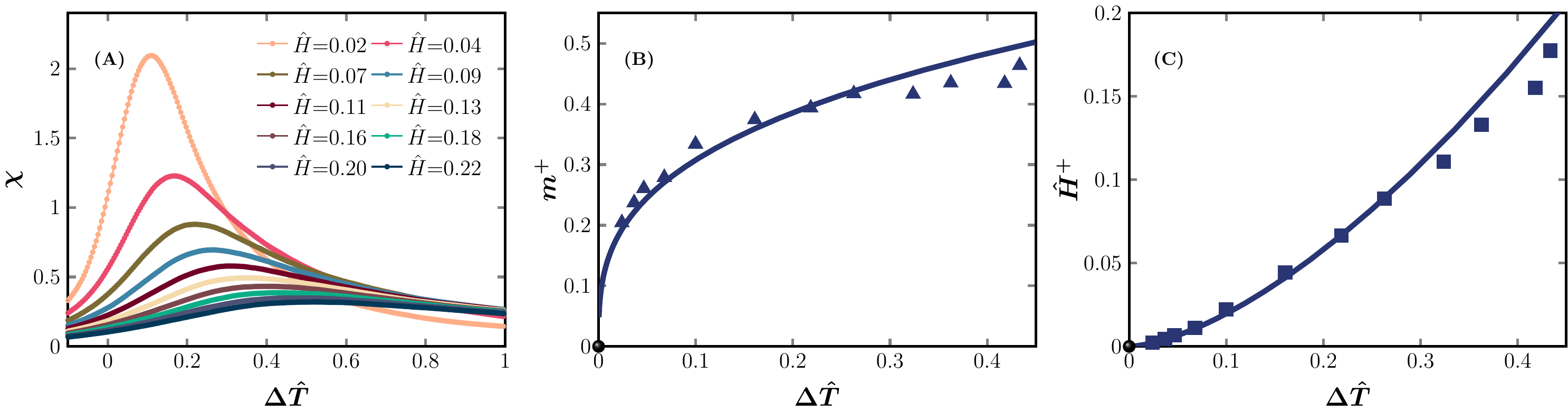}
  \caption{{\bf Supercritical crossover lines $L^{\pm}$ in the 3D Ising model.} 
  %\YJ{$\hat{H}$ not $\hat{\rm H}$, check everywhere}
  Results are obtained from Monte Carlo simulations.
(A) Magnetic susceptibility $\chi$ as a function of the reduced temperature $\hat{T}$ (for this panel the lattice size is $12 \times 12 \times 12$), for a few different reduced magnetic fields $\hat{H}$. The peaks determine the supercritical crossover line $L^{+}$. 
(B) The magnetization $m^{+}$ and (C) reduced field $\hat{H}^+$ along the $L^{+}$ line are plotted as functions of 
$\Delta \hat{T}$. 
The lines represent fits, $m^+ = A_m \Delta \hat{T} ^\beta$ and $\hat{H}^+ = A_H \Delta \hat{T} ^{\beta+\gamma}$, where $A_m = 0.652$ and $A_H= 0.719$ are fitting parameters (see Fig.~\ref{fig:Ising_MC_3D_log} for corresponding log-log plots).
%\YJ{Is it possible to add one or two points close to the critical point?}}
  }
  \label{fig:Ising_MC_3D}
\end{figure}

\begin{figure}[!htbp]
  \centering
  \includegraphics[width=0.7\linewidth]{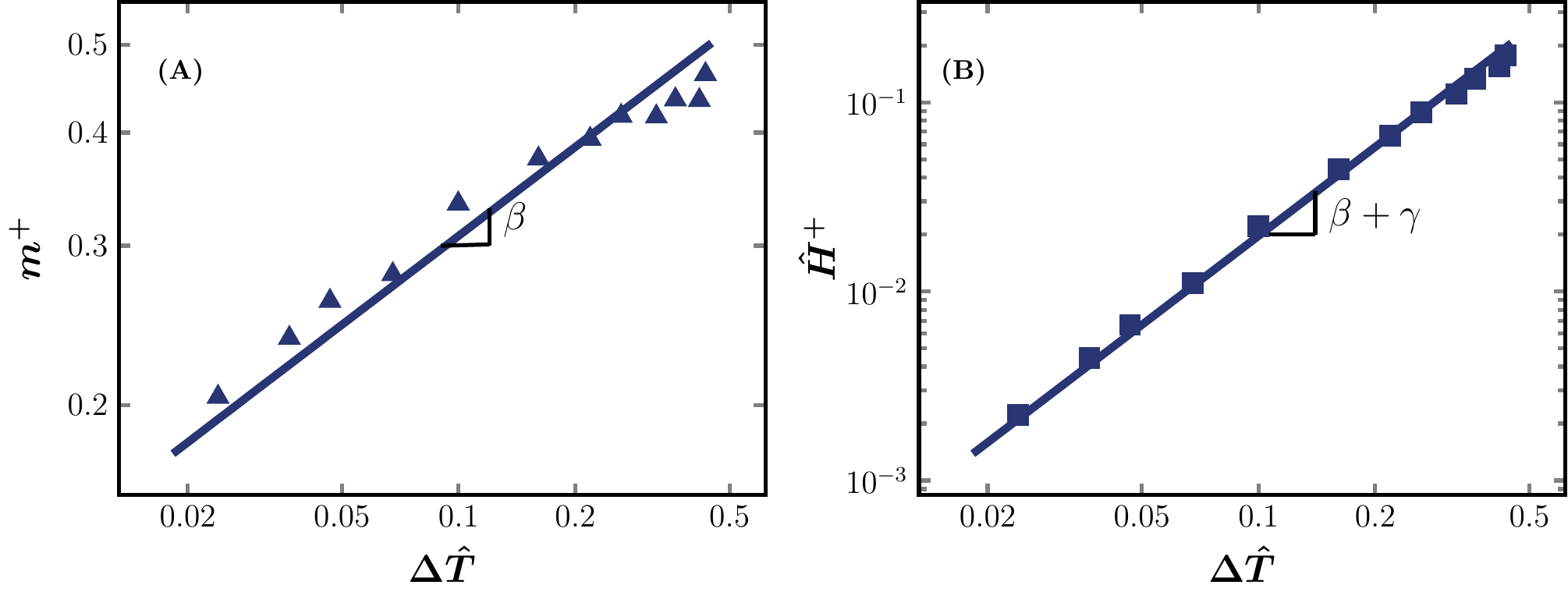}
  \caption{{\bf  Scaling laws  of the $L^+$ line in the 3D Ising model.} 
  (A) The magnetization $m^{+}$ and (B) reduced field $\hat{H}^+$  as functions of $\Delta \hat{T}$, in log-log scales. 
The lines represent fits, $m^+ = A_m \Delta \hat{T} ^\beta$ and $\hat{H}^+ = A_H \Delta \hat{T} ^{\beta+\gamma}$, where $A_m = 0.652$ and $A_H= 0.719$ are fitting parameters. 
 % \blue{{\bf Scaling law verification of 3D Ising model.} (A) $L^+$ of 3D Ising model in log-log scaled $m-\Delta{\hat{T}}$ phase diagram. The points are fitting by $f(x) = \beta x+b$, where $b=-0.50$. (B) $L^+$ of 3D Ising model in log-log scaled $\hat{H}-\Delta{\hat{T}}$ phase diagram. The points are fitting by $f(x) =(\beta+\gamma) x+b$, where $b=-0.43$. Solid lines are fitting curves.}
  }
  \label{fig:Ising_MC_3D_log}
\end{figure}

We study the Ising model in both two (2D) and three (3D) dimensions to validate the scaling relationships Eqs.~(1) and~(2), using Monte Carlo simulations with the wolff algorithm~\cite{wolff1989collective, newman1999monte}. 
%\YJ{add citation of the wolff algorithm} 
The lattice size is $40 \times 40$ in 2D and $25 \times 25 \times 25$ in 3D. The mean magnetization is computed using the definition,
$
m = \frac{1}{N} \left \langle \sum_{i=1}^N s_i \right \rangle$,
and the magnetic susceptibility is computed using,
\beq
\chi = \left( \frac{\partial m}{\partial H} \right)_{T} = \beta N \left( \left \langle m^2 \right \rangle - {\left \langle m \right \rangle}^2 \right).
\eeq
%where $\langle \cdots \rangle$ represents the ensemble average.
%\YJ{how did you calculate $\chi$ in simulations? first equality or the second equality?}
In simulations, $m$ and $\chi$ are obtained by averaging over $10^4$ independent equilibrium configurations.
%We take 1000 equilibrium configurations to approximate ensemble average. Besides, we conduct 10 seperate simulations and take their average value to reduce fluctuation. Savgol filter (a function in python) also has been used to smooth the $\chi-\Delta\hat{T}$ curve. 
For a few given $\hat{H}$, the magnetic susceptibility $\chi$ is plotted as a function of temperature $\hat{T}$ in Fig.~\ref{fig:Ising_MC_2D}(A) and Fig.~\ref{fig:Ising_MC_3D}(A), from which the peaks are located. 
The supercritical crossover lines determined in this way can be well fitted to scalings Eqs.~(1) and~(2), as shown in Figs.~\ref{fig:Ising_MC_2D}-\ref{fig:Ising_MC_3D_log}.
In the fitting, the exponents are fixed to the known values: $\beta = 1/8$, $\gamma = 7/4$ in 2D~\cite{fisher1967theory}, and $\beta = 0.326$, $\gamma = 1.237$ in 3D~\cite{guida1998critical};  the prefactor is treated as a fitting parameter.
%some different given magnetic fields $H$ varies from 0.1 to 1.2, we calculate the magnetic susceptibility $\chi$ as a function of temperature $\hat{T}$, then obtain $\hat{T}$ and $m$ that correspond to the maximal $\chi$. Therefore, we can get the $\Delta \hat{T} - \hat{H}$ and $\Delta \hat{T} - m$ phase diagrams when $\chi$ is maximized, as can be seen in Fig.~\ref{fig:Ising_MC_2D} and ~\ref{fig:Ising_MC_3D}. They fall on $f(x) = a x^{\beta+\gamma}$ and $f(x) = a x^\beta$ respectively, where a is a fitting parameter, and for 2D model $\beta = 1/8$, $\gamma = 7/4$~\cite{fisher1967theory}, for 3D model $\beta = 0.326$, $\gamma = 1.237$~\cite{guida1998critical}. Their logarithmic curves further verify the scalings Eqs.~(\ref{eq:scaling1}) and~(\ref{eq:scaling2}), as shown in Fig.~\ref{fig:Ising_MC_2D_log} and ~\ref{fig:Ising_MC_3D_log}.
%}  
%\clearpage

\section{Summary of supercritical crossover lines in the literature}

Table~\ref{table:lines} summarizes supercritical  crossover lines existing in the literature, compared to the proposal of this study.

\begin{table*}[!htbp]
\centering
\caption{Summary of crossover lines in supercritical fluids.}

\begin{tabular}{lrrr}
crossover lines & number of lines & emanating from the critical point? & obeying critical scalings Eq.(3) and (4)? \\
\midrule
1. Widom line~\cite{xu2005relation}  &  one & yes & no   \\
2. Fisher-Widom line~\cite{fisher1969decay}  &  one & no & no   \\
3. Frenkel line~\cite{brazhkin2012two}  &  one & no & no   \\
4. Nishikawa line~\cite{nishikawa1995correlation}  &  one & yes & no   \\
5. symmetry line~\cite{ploetz2019gas}  &  one & yes & no   \\
6. percolation lines~\cite{woodcock2013observations}  &  two & no & no   \\
7. pseudo-boiling boundaries~\cite{banuti2015crossing}  &  two & yes & no   \\
8. $L^\pm$ lines (this study) &  two & yes & yes   \\
\bottomrule
\end{tabular}
\label{table:lines}
\end{table*}

\section{van der Waals equation of state}
\label{sec:vdw}

\begin{figure*}[!htbp]
  \centering
  \includegraphics[width=\linewidth]{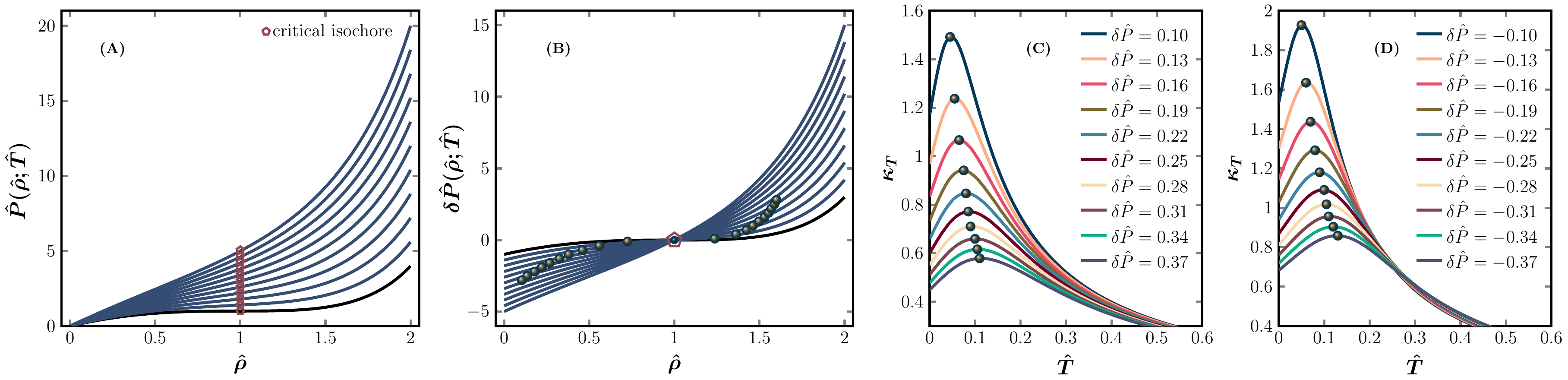}
  \caption{{\bf Determination of crossover lines $L^{\pm}$ of the van der Waals equation of state}.
  (A) Isothermal EOSs, $\hat{P}(\hat{\rho}; \hat{T})$, above the critical temperature. The black line is critical isotherm, and dark blue lines are other isotherms (from bottom to top, $\hat{T} = 1.00, 1.10, 1.20, 1.30, 1.40, 1.50, 1.60, 1.70, 1.80, 1.90,  2.00.$).
  The critical isochore is marked by pentagons. (B) Shifted EOSs $\delta \hat{P}(\hat{\rho}; \hat{T})$.
  %, for which the pressure $\hat{P}(1; \hat{T})$ along the  critical isochore is subtracted from each EOS. 
  The Susceptibility 
 $\kappa_T 
    = \left( \frac{\partial \hat{\rho}}{\partial  \hat{P}} \right)_{\hat{T}}$ is plotted as a function of $\hat{T}$, for a few fixed (C) $\delta \hat{P}>0$ and (D) $\delta \hat{P}<0$; the loci of peaks determine $L^+$ and $L^-$ respectively (circles). The meanings of symbols are the same in all panels. %\YJ{$\delta \hat{P}$, $\kappa_T$}
 }
  \label{fig:VDF}
\end{figure*}

%Just as the Ising model is a simplified model of the ferromagnetic phase transition, 
The van der Waals equation is one of the most widely used equation of state (EOS) 
 for describing the liquid-gas phase transition. 
 %Under the dimensionless condition, 
 The reduced form of the van der Waals equation is,
\beq
\left(\hat{P}+\frac{3}{\hat{V}^2} \right) \left(3\hat{V}-1 \right) = 8\hat{T},
\label{eq:van der Waals equation}
\eeq 
where $\hat{P} = P/P_{\rm c}$, $\hat{V}=V/V_{\rm c}$ and $\hat{T} = T/T_{\rm c
}$ are the reduced pressure, volume and temperature respectively, and $P_{\rm c}$, $V_{\rm c}$ and $T_{\rm c}$  are the pressure, volume and temperature of the critical point.

%There are no parameters related to material properties in the equation, so the equation is universal.
%\YJ{add one figure of four panels, similar to Fig. 2 A, B, C D.}
\subsection{Determination  of $L^\pm$ lines}
The supercritical crossover lines $L^{\pm}$ are obtained numerically following the procedure described below. A similar procedure is used to obtain $L^{\pm}$ of argon in the main text (see Fig. 2).
\begin{itemize}
    \item Plot the isothermal EOSs $\hat{P}(\hat{\rho}; \hat{T})$ above the critical temperature, where $\hat{\rho} = 1/\hat{V}$ is the reduced density (see Fig.~\ref{fig:VDF}(A)).
    \item Subtract the critical isochore $\hat{P}(1; \hat{T})$ from the original EOS $\hat{P}(\hat{\rho}; \hat{T})$. In the shifted EOSs,  $\delta \hat{P}(\hat{\rho}; \hat{T}) = \hat{P}(\hat{\rho}; \hat{T}) - \hat{P}(1; \hat{T}) $, the critical isochore line $\hat{P}(1; \hat{T})$  becomes a single point $\delta \hat{P}(1; \hat{T}) = 0$ that coincides with the critical point  (see Fig.~\ref{fig:VDF}(B)). All shifted EOSs $\delta \hat{P}(\hat{\rho}; \hat{T})$ intersect at this point.
    \item Along constant-$\delta \hat{P}$ paths, find the maxima of the susceptibility $\kappa_T = \left( \frac{\partial \hat{\rho}}{\partial (\delta \hat{P})} \right)_{\hat{T}}$, which give $L^{\pm}$ (see Fig.~\ref{fig:VDF}(C) and (D)). The constant-$\delta \hat{P}$ paths correspond to  lines parallel to the the critical isochore in the $\hat{P}-\hat{T}$ phase diagram.
   % \YJ{unify the notation of $\kappa_T = \left( \frac{\partial \hat{\rho}}{\partial (\delta \hat{P})} \right)_{\hat{T}}$ and $\kappa_T = \left ( \frac{\partial \hat{\rho}}{\partial (\delta \hat{P})} \right ){\hat{T}}$  in the paper.}
\end{itemize}

\subsection{$L^{\pm}$ lines in pressure-temperature and pressure-volume phase diagrams}

\begin{figure*}[!htbp]
  \centering
  \includegraphics[width=\linewidth]{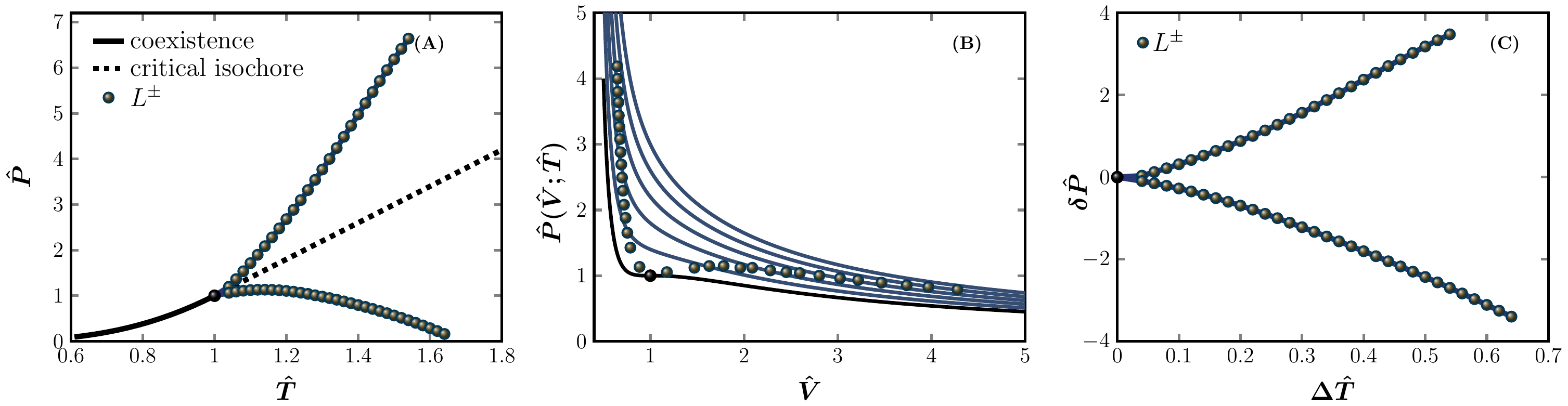}
  \caption{{\bf Crossover lines $L^{\pm}$ of the van der Waals equation of state.} 
  (A)  Pressure-temperature ($\hat{P}-\hat{T}$) diagram. (B) Pressure-volume ($\hat{P}-\hat{V}$)  diagram, where the lines are isotherms for $\hat{T} = 1.00, 1.10, 1.20, 1.30, 1.40, 1.50.$ (from bottom to top). (C)  $\delta \hat{P}-\Delta \hat{T}$  diagram. %\YJ{$L^{\pm}$}
%  \YJ{unify the symbols for $L^{\pm}$ in three panels.}
 }
  \label{fig:VDF_PVT}
\end{figure*}
The above determined crossover lines $L^{\pm}$ are plotted in pressure-temperature ($\hat{P}-\hat{T}$) and pressure-volume ($\hat{P}-\hat{V}$) diagrams (see Fig. ~\ref{fig:VDF_PVT}(A) and Fig.~\ref{fig:VDF_PVT}(B)).
%\YJ{Make the figure. Check if it makes sense to study the  critical scaling of $L^{\pm}$ in the  van der Waals equation.} 
In the $\delta \hat{P}-\Delta \hat{T}$  diagram, $L^{+}$ is nearly symmetric to $L^{-}$  (see Fig.~\ref{fig:VDF_PVT}(C)), similar to the case of Ising model (see  Fig.~\ref{fig:Ising}A).
%for the Ising model.

%According to definition of widom line~\cite{xu2005relation}, we calculate the compressibility given by the van der Waals model on the path of fixed pressures (the dashed line in Fig.~\ref{fig:VW}(B)), and obtain the maximum value. The extreme points corresponding to different pressures are combined into the widom line in P-T phase diagram, as shown in Fig.~\ref{fig:VW}(B). Obviously, this diagram is completely different from the phase diagram of the Ising model shown in Figure Fig.~\ref{fig:VW}(A). Basically, the number of lines is different. However, as stated in the maintext, if we choose the critical isochore as a reference line, then take its parallel lines (dashed lines in Fig.~\ref{fig:VW}(C)) as observation paths to calculate the maximum value of compressibility, the $O(1)$ critical universality of ferromagnetic phase transition and liquid-gas phase transition will be satisfied again. In other words, such a calculation gives two extreme lines $L^{\pm}$, which are approximately symmetric with respect to the critical isochore, and they satisfy the scaling relation $ |\delta \hat{P}|  \sim (T - T_{\rm c})^{\beta + \gamma}$.

\section{$L^\pm$ lines of  water}
\begin{figure}[!htbp]
  \centering
 \includegraphics[width=\linewidth]{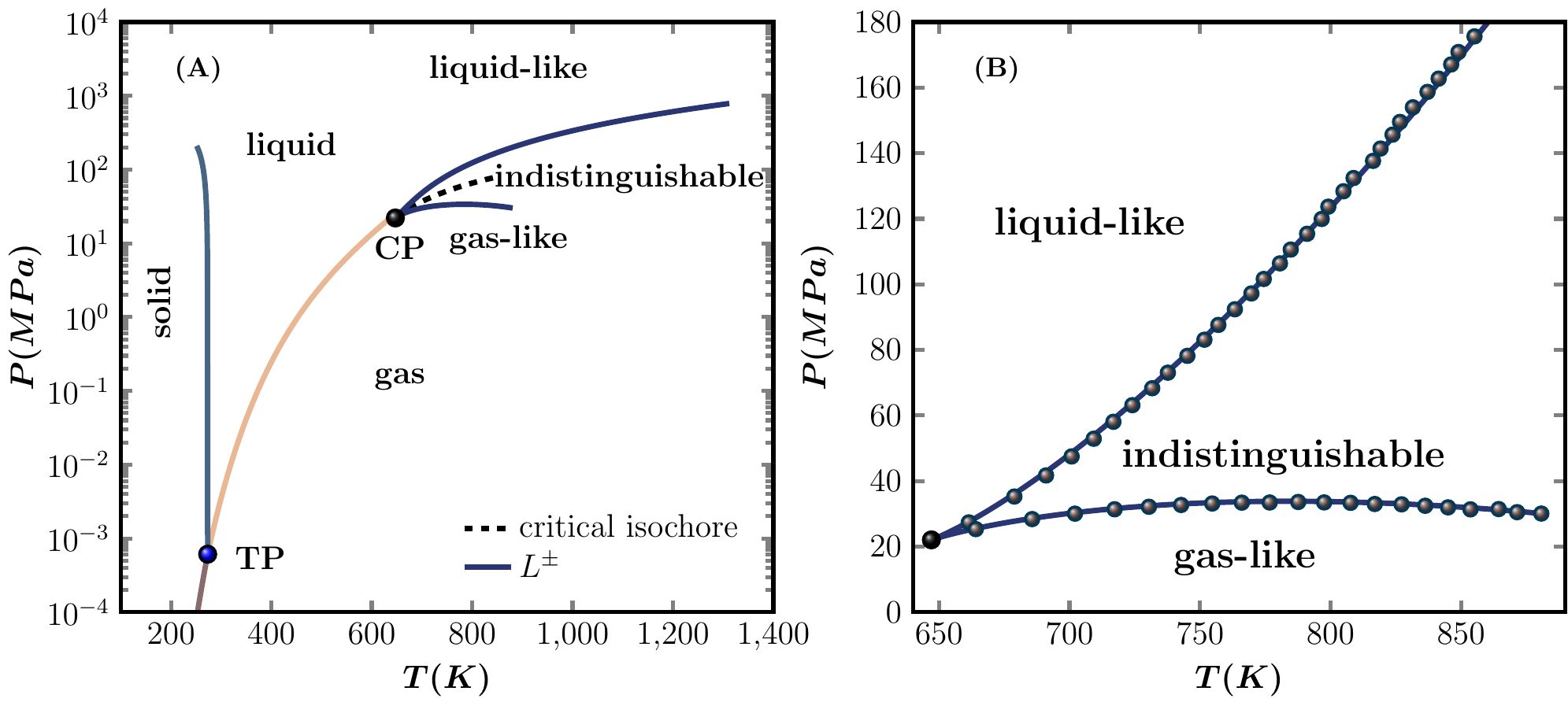}
  \caption{ {\bf Supercritical crossover  lines $L^\pm$ of water.} (A)  Pressure-temperature phase diagram of water. 
  %The blue, brown and light orange line are coexistence lines of solid-liquid, solid-gas and gas-liquid phase transition respectively. 
  $L^{\pm}$ lines have been included in the diagram, separating liquid-like, indistinguishable and gas-like  regimes.
  The triple point (TP) and critical point (CP) are also indicated. 
  (B)  $L^\pm$ lines in the vicinity of the critical point. Points are determined from the maximum of $\kappa_T$. Lines represent fitting to the data points, following the second fitting method explained in Sec.~\ref{sec:fitting}.
  %\YJ{$L^\pm$}
  %\YJ{Try to use consistent colors/symbols for $L^{\pm}$ in all figures. Remove $L^{\pm}$ points in (A) and keep only lines; the points are in (B). You can "replace triple" point by TP, and "critical point" by CP to have same space. Add isochore line in (A), to show that $L^+$ is not a simple extension of coexistence line. All words should be in black color, and as large as possible. In (A), the y-axis can start from, e.g, $10^{-4}$. The fitting to $L-$ shall be improved: the rightmost points do not fall on the line. For $L^-$, the fitting line can be stopped at the last data point; don't extend it below the critical point.}
  %\YJ{use the same blue color for $L-$ in A}
  %in the supercritical region.
  }
  \label{fig:water}
\end{figure}

%Show the phase diagram of water with $L^+$ and $L^-$. Water is important. This will be useful for other people. 
%Water is one of the most common and important substances in nature, and 

Crossovers lines of supercritical water have been investigated in many studies~\cite{gallo2016water,gallo2014widom,corradini2015widom, mishima1998relationship,smart2018war,kim2017maxima, strong2018percolation,matsugami2014theoretical,fomin2015dynamical}, using various definitions  listed in Table~\ref{table:lines}. Figure~\ref{fig:water}(A) shows the phase diagram of water~\cite{chaplin2019structure}, %\YJ{replace the citation by papers or books},
where we have included   thermodynamic crossover lines $L^{\pm}$.
The original EOS data are collected from the NIST database~\cite{NIST}, and $L^{\pm}$ are determined in the same way as for argon (see Fig.~\ref{fig:water}(B) for an enlarged view). 
%\YJ{Make two panels. (A) log-linear plot. Extrapolate $L^+$ using Eq.~(\ref{eq:fit_Lp}); extend the phase diagram to the range comparable to Fig. 3. (B) linear-linear plot with an enlarged view near the critical point. }

%In the same way as the calculation of Ar, we collect the isotherm data of water from NIST database, and calculate the extreme value of susceptibility $\kappa_T \equiv \left( \frac{\partial \hat{\rho}}{\partial (\delta \hat{P})} \right)_{\hat{T}}    = \left( \frac{\partial \hat{\rho}}{\partial  \hat{P}} \right)_{\hat{T}}
%    =  \left( \frac{P_{\rm c}}{\rho_{\rm c}}  \frac{\partial \rho}{\partial  P} \right)_T$ according to the method described in the maintext, thus giving two dynamic crossover lines $L^{\pm}$, and finally convert them into the original $P-T$ phase diagram, with the dimensions of pressure and temperature being restored.

%\section{Linear scaling theory}
\section{Linear scaling theory}
\label{sec:linear scaling theory}

\subsection{Sketch of the theory}
Near a critical point, the themodynamic potential $\Psi(h_1,h_2)$ is a homogeneous function,
\beq
\Psi(h_1,h_2) = |h_2|^{2-\alpha} f\left(h_1/|h_2|^{\beta+\gamma}  \right),
\label{eq:Phi}
\eeq
of two independent scaling fields $h_1$ and $h_2$~\cite{widom1965equation, widom1974critical, anisimov1998nature, fuentevilla2006scaled}. Here $h_1$ and $h_2$ are the {\it ordering field} and {\it thermal field}, $f(x)$ is a universal scaling function, and $\alpha$, $\beta$ and $\gamma$ are universal critical exponents interrelated by the scaling relationship $\alpha + 2 \beta + \gamma = 2$.
We follow Ref.~\cite{luo2014behavior} and express $h_1$ and $h_2$ as linear 
 combinations of physical fields $P$ and $T$,
\beq
\begin{aligned}
h_1 = \Delta\hat{P}\cos{\varphi}-\Delta\hat{T}\sin{\varphi},\\
h_2 = \Delta\hat{P}\sin{\varphi}+\Delta\hat{T}\cos{\varphi}.
\label{eq:h_12_PT}
\end{aligned}
\eeq
Here $\Delta\hat{P} = P/P_{\rm c}-1$, $\Delta\hat{T} = T/T_{\rm c}-1$, and $\tan\varphi =  d\hat{P}/d\hat{T}$ is the slope of the coexistence line. For {\it symmetric} systems (e.g., the Ising model and the lattice-gas model), $\varphi =0$: in this case the ordering field $h_1 = \Delta\hat{P}$ and the thermal field $h_2 = \Delta\hat{T}$. Real liquid-gas transitions are generally considered as {\it asymmetric}, because  $\varphi \neq 0$; consequently the physical fields $P$ and $T$ are mixed of $h_1$ and $h_2$.

For practical use, it is convenient to employ a parametric representation of $h_1$ and $h_2$. The linear scaling
theory~\cite{schofield1969parametric,schofield1969correlation} provides the simplest form of  parameterization, which expresses $h_1$ and $h_2$ with ``polar'' variables $r$ and $\theta$:
\beq
\begin{aligned}
h_1  &= ar^{\beta\delta}\theta(1-\theta^2),\\
h_2  &= r(1-b^2\theta^2),
\label{eq:h_12}
\end{aligned}
\eeq
where $\delta$ is an independent exponent satisfying the scaling relation $(\delta -1) \beta = \gamma$,  $a$ is a system-dependent fitting parameter, and $b$ is a universal parameter obeying  $b^2 = (\delta-3)/(\delta-1)(1-2\beta)$. The variable $r$ represents a ``distance'' from the critical point, and $\theta\in[-1,1]$ is a measure of the ``angle'' from the coexistence line (the coexistence line corresponds to $\theta = \pm 1$).
%taking values of -1 and +1 on different sides of the coexistence line.
With the parameterization Eq.~(\ref{eq:h_12}), Eq.~(\ref{eq:Phi}) becomes,
%If one assume that the themodynamic potential $\Phi(h_1,h_2)$ has the form
\beq
\Psi(r,\theta) = r^{\beta(\delta+1)}p(\theta),
\label{eq:themodynamic potential}
\eeq
where $p(\theta)$ is an analytical function.

The order parameter conjugated to $h_1$ is 
\beq
\begin{aligned}
\phi_1 &= \left(\frac{\partial\Psi}{\partial h_1}\right)_{h_2} = \frac{\partial\Psi}{\partial r} \frac{\partial r}{\partial h_1} + \frac{\partial\Psi}{\partial\theta} \frac{\partial \theta}{\partial h_1} \\
       &=r^\beta m(\theta)\\
        & = k r^\beta \theta,
\label{eq:order parameter1}
\end{aligned}
\eeq
where $m(\theta)$ is related to $p(\theta)$ in Eq.~(\ref{eq:themodynamic potential}), which is in general unknown. 
In the last equality,  $m (\theta)$ is assumed to be a linear function of $\theta$, $m(\theta) =k\theta$, hence the name linear scaling theory~\cite{schofield1969parametric,schofield1969correlation}. Here $k$ is another system-dependent fitting parameter. Similarly, the order parameter conjugated to $h_2$ can be calculated,
\beq
\begin{aligned}
\phi_2 &= \left(\frac{\partial\Psi}{\partial h_2}\right)_{h_1} = \frac{\partial\Psi}{\partial r} \frac{\partial r}{\partial h_2} + \frac{\partial\Psi}{\partial\theta} \frac{\partial \theta}{\partial h_2} \\
       &=r^{\beta(\delta+1)-1} s(\theta).
\label{eq:order parameter2}
\end{aligned}
\eeq
The function $s(\theta)$ can be determined from the equality $\left(\frac{\partial\phi_1}{\partial h_2}\right)_{h_1}=\left(\frac{\partial\phi_2}{\partial h_1}\right)_{h_2}$, giving, 
\beq
s(\theta) &=a k(s_0+s_2\theta^2), 
\eeq
where,
\beq
%\resizebox{0.9\hsize}{!}
%{$
\begin{aligned}
s_0 &= -{\frac{\beta\left[-3+\delta+b^2(-1+\delta)(-2+\beta+\beta\delta)\right]}{2b^4(-2+\beta+\beta\delta)(-1+\beta+\beta\delta)}},\\
s_2 &= \frac{\beta(\delta-3)}{2b^2(-2+\beta+\beta\delta)}.
\label{eq:order parameter2 coefficient}
\end{aligned}
%$}
\eeq

Let us summarize the input to the theory: the critical exponents ($\beta, \gamma, \ldots$) determined by the universality class, and three system-dependent parameters including the slope of the coexistence line $\varphi$, the parameter $a$ appeared in Eq.~(\ref{eq:h_12}), and the parameter $k$ appeared in Eq.~(\ref{eq:order parameter1}). Once these values are given, physical quantities depend only on two independent variables $r$ and $\theta$, and
the EOSs are fixed. 
For example, using  Eqs.~(\ref{eq:h_12_PT}) and~(\ref{eq:h_12}),  $\Delta\hat{P}$ and $\Delta\hat{T}$ can be expressed as,
\beq
\begin{aligned}
\Delta\hat{P} &= h_1\cos\varphi+h_2\sin\varphi\\
&= ar^{\beta+\gamma}\theta(1-\theta^2)\cos\varphi + r(1-b^2\theta^2)\sin\varphi,\\
\Delta\hat{T} &= h_2\cos\varphi-h_1\sin\varphi\\
&= r(1-b^2\theta^2)\cos\varphi - ar^{\beta+\gamma}\theta(1-\theta^2)\sin\varphi.
\label{eq:physical fields}
\end{aligned}
\eeq
The reduced volume is
\beq
\begin{aligned}
\Delta\hat{V} &= \left( \frac{\partial\Psi}{\partial \Delta\hat{P}} \right)_{\hat{T}}
       = \frac{\partial\Psi}{\partial h_1} \frac{\partial h_1}{\partial \Delta\hat{P}} + \frac{\partial\Psi}{\partial h_2} \frac{\partial h_2}{\partial \Delta\hat{P}} \\
       &=-\phi_1\cos\varphi-\phi_2\sin\varphi,
\label{eq:volume}
\end{aligned}
\eeq
Near the critical point, the reduced density is $\Delta\hat{\rho} = (\rho-\rho_{\rm c})/\rho_{\rm c} \simeq - \Delta\hat{V}$,
%and we can neglect the difference between $-\Delta\hat{V} = -(V-V_c)/V_c=(\rho-\rho_{\rm c})/\rho$ and $(\rho-\rho_{\rm c})/\rho_{\rm c}$ near the critical point, 
and therefore 
\beq
\Delta\hat{\rho} = \phi_1\cos\varphi+\phi_2\sin\varphi.
\label{eq:density}
\eeq
Similarly, the reduced entropy is
\beq
\begin{aligned}
\Delta\hat{S} &= -\left(\frac{\partial\Psi}{\partial \Delta\hat{T}} \right)_{\hat{P}}
       = -\left(\frac{\partial\Psi}{\partial h_1} \frac{\partial h_1}{\partial \Delta\hat{T}} + \frac{\partial\Psi}{\partial h_2} \frac{\partial h_2}{\partial \Delta\hat{T}}\right) \\
       &=-\phi_1\sin\varphi+\phi_2\cos\varphi.
\end{aligned}
\label{eq:entropy}
\eeq
The susceptibilities can be computed using the following expressions,
\beq
\begin{aligned}
&\chi_1 \equiv 
\left(\frac{\partial\phi_1}{\partial h_1}\right)_{h_2}= \frac{k}{a}r^{-\gamma}C_1(\theta),\\
&\chi_2 \equiv \left(\frac{\partial\phi_2}{\partial h_2}\right)_{h_1}= r^{-\alpha}C_2(\theta),\\
&\chi_{12} \equiv \left(\frac{\partial\phi_1}{\partial h_2}\right)_{h_1}=\left(\frac{\partial\phi_2}{\partial h_1}\right)_{h_2}= k r^{\beta-1}C_{12}(\theta),
\label{eq:Response function}
\end{aligned}
\eeq
where
\beq
%\resizebox{0.88\hsize}{!}
%{
\begin{aligned}
&C_1(\theta)=\frac{(1-b^2\theta^2(1-2\beta))}{C_0(\theta)},\\
&C_2(\theta)=\frac{\left[(1-\alpha)(1-3\theta^2)s(\theta)-2s_2\beta\delta\theta^2(1-\theta^2)\right]}{C_0(\theta)},\\
&C_{12}(\theta)=\frac{\beta\theta\left[1-\delta-\theta^2(3-\delta)\right]}{C_0(\theta)},\\
&C_0(\theta)=(1-3\theta^2)(1-b^2\theta^2)+2\beta\delta b^2\theta^2(1-\theta^2).
\label{eq:Response function coefficient}
\end{aligned}
%$}
\eeq
Depending on $\varphi$, the physical response functions are generally combinations of $\chi_1$, $\chi_2$ and $\chi_{12}$. For example, the susceptibility (isothermal compressibility) $\kappa_T$, the isobaric specific heat $C_{P}$, and the isobaric thermal expansion coefficient $\alpha_{P}$ are, 
%\YJ{I think it is better to use notations  $\kappa_T$, $C_{P}$ and $\alpha_{P}$. Keep them consistent everywhere. Be careful about the lower and upper cases, e.g., $C_P$ vs $C_P$. Replace $\kappa_T$ by $\kappa_T$, $\beta$ by $\beta_T$.}
\beq
\begin{aligned}
 &\kappa_T \equiv 
 %\left( \frac{\partial \hat{\rho}}{\partial \hat{P}} \right)_{\hat{T}}
  \left(  \frac{\partial \Delta \hat{\rho}}{\partial \Delta \hat{P}}\right)_{\hat{T}}
 =  \left( \frac{\partial \Delta \hat{\rho}}{\partial h_1} \frac{\partial h_1}{\partial \Delta \hat{P}} +  \frac{\partial \Delta \hat{\rho}}{\partial h_2} \frac{\partial h_2}{\partial \Delta \hat{P}} \right)
 =  \left(\chi_1 \cos^2{\varphi} + \chi_{12} \sin{2\varphi} + \chi_2 \sin^2{\varphi} \right),\\
 &C_P \equiv 
% \left( \frac{\partial \hat{S}}{\partial \hat{T}} \right)_{\hat{P}}= 
\left( \frac{\partial \Delta \hat{S}}{\partial \Delta \hat{T}} \right)_{\hat{P}}
 =  \left( \frac{\partial \Delta \hat{S}}{\partial h_1} \frac{\partial h_1}{\partial \Delta \hat{T}} +  \frac{\partial \Delta \hat{S}}{\partial h_2} \frac{\partial h_2}{\partial \Delta \hat{T}} \right)
 = \left(\chi_1 \sin^2{\varphi} - \chi_{12} \sin{2\varphi} + \chi_2 \cos^2{\varphi} \right),\\
  &\alpha_P \equiv 
  %\left( \frac{\partial \hat{V}}{\partial \hat{T}} \right)_{\hat{P}}
   \left(  \frac{\partial \Delta \hat{V}}{\partial \Delta \hat{T}} \right)_{\hat{P}}
 = \left( \frac{\partial \Delta \hat{V}}{\partial h_1} \frac{\partial h_1}{\partial \Delta \hat{T}} +  \frac{\partial \Delta \hat{V}}{\partial h_2} \frac{\partial h_2}{\partial \Delta \hat{T}} \right)
 =  \frac{1}{2}\left[(\chi_1-\chi_2) \sin{2\varphi} -2 \chi_{12} \cos{2\varphi} \right].
 \label{eq:response}
\end{aligned}
\eeq
Below we show that, independent of $\varphi$, the pressure and order parameter $\Delta \hat{\rho}$ on the supercritical crossover lines satisfy the following scalings (identical to Eqs.~(3) and~(4)),
\beq
|\delta \hat{P}^{\pm}|  \sim \Delta \hat{T}^{\beta + \gamma},
\label{eq:scaling_P}
\eeq
where $\delta \hat{P}  = \Delta \hat{P}(\Delta \hat{\rho}; \hat{T}) - \Delta \hat{P}(0; \hat{T})$ with $\Delta \hat{P}(0; \hat{T})$ the reduced pressure on the critical isochore,
and 
\beq
|\Delta \hat{\rho}^{\pm}|  \sim \Delta\hat{T}^{\beta}.
\label{eq:scaling_rho}
\eeq
Note that on the critical isochore, the reduced density is $\Delta \hat{\rho} = 0$.

%whose explicit form is not essential in the following derivation. 

\subsection{Symmetric model with $\varphi = 0$ }
When the slope of the coexistence line is zero, i.e., $\varphi = 0$, the scalings Eqs.~(\ref{eq:scaling_P}) and~(\ref{eq:scaling_rho})
%Eq.~(3) and (4) 
can be derived analytically~\cite{luo2014behavior}.
%\YJ{Replace $\delta$ by $\beta$, $\gamma$ in the following repression.}
In this case,
\beq
\begin{aligned}
&\Delta \hat{P} = h_1  = ar^{\beta+\gamma}\theta(1-\theta^2),\\
&\Delta \hat{T} = h_2  = r(1-b^2\theta^2),\\
&\Delta \hat{\rho} = \phi_1 = k r^\beta \theta,
\end{aligned}
\label{eq:PTrho_phi0}
\eeq
from which we can get
%\beq
%\begin{aligned}
$r  = \left[\frac{\Delta \hat{P}}{a\theta(1-\theta^2)}\right]^{1/(\beta+\gamma)}$.
The response functions Eqs.~(\ref{eq:response}) become,  
%= (\frac{\Delta \hat{\rho}}{k \theta})^{1/\beta},\\
%\end{aligned}
%\label{eq:r_phi0}
%\eeq
%therefore, 
%\beq
%\Delta \hat{T}  = \left(\frac{\Delta \hat{P}}{a\theta(1-\theta^2)}\right)^{1/(\beta+\gamma)}(1-b^2\theta^2) = \left(\frac{\Delta \hat{\rho}}{k \theta}\right)^{1/\beta}(1-b^2\theta^2),
%\label{eq:T_phi0}
%\eeq
\beq
\kappa_T = \frac{k}{a}r^{-\gamma}C_1(\theta) =   \frac{k}{a} \left[\frac{a\theta(1-\theta^2)}{\Delta \hat{P}}\right]^{\gamma/(\beta+\gamma)}C_1(\theta).
\label{eq:compressibility_phi0}
\eeq
\beq
C_P =  r^{-\alpha}C_2(\theta) =   \left[\frac{a\theta(1-\theta^2)}{\Delta \hat{P}}\right]^{\alpha/(\beta+\gamma)}C_2(\theta),
\label{eq:heat_capacity_phi0}
\eeq
\beq
\alpha_P = - k r^{\beta-1}C_{12}(\theta) =  - k  \left[\frac{a\theta(1-\theta^2)}{\Delta \hat{P}}\right]^{(1-\beta)/(\beta+\gamma)}C_{12}(\theta).
\label{eq:Isobaric_expansion_phi0}
\eeq
For a fixed $\Delta \hat{P}$, a response function reaches its extreme value at  a constant $\theta$: $\theta_1 = \pm0.526$, $\theta_2 = \pm0.925$ and  $\theta_{12} = \pm0.743$ for $\kappa_T$, $C_P$ and $\alpha_P$ respectively.
%, which are constants independent of $\Delta \hat{P}$.
For any response function, because its  extreme value  is  obtained at a constant  $\theta$, from Eq.~(\ref{eq:PTrho_phi0}) we can derive universal  scalings (\ref{eq:scaling_P}) and
Eqs.~(\ref{eq:scaling_rho}), along the line of extremums. Note that, when $\varphi = 0$, $\delta \hat{P} = \Delta \hat{P} - \Delta \hat{P}(\Delta \hat{\rho} = 0; \hat{T}) = \Delta \hat{P}$, since the isochore line is identical to the horizontal axis $\Delta \hat{P}(0; \hat{T}) = 0$.

%$\Delta \hat{P}(0; \hat{T}) = 0$ at $\varphi = 0$, so $\delta \hat{P}  = \Delta \hat{P}$.

%relationships ,
%\beq
%\Delta \hat{T}  \sim \Delta \hat{P}^{1/(\beta+\gamma)},
%\label{eq:TP_scaling}
%\eeq
%\beq
%\Delta \hat{T} \sim \Delta \hat{\rho}^{1/\beta},
%\label{eq:Trho_scaling}
%\eeq
%which are identical to Eqs.~(\ref{eq:scaling_rho}) and (\ref{eq:scaling_P}), 
%and are the same with all response functions. Besides, 
%$\Delta \hat{P}(0; \hat{T}) = 0$ at $\varphi = 0$, so %$\delta \hat{P}  = \Delta \hat{P}$.
%}
\subsection{Asymmetric models with $\varphi \neq 0$ }

\begin{figure}[!htbp]
  \centering
  \includegraphics[width=\linewidth]{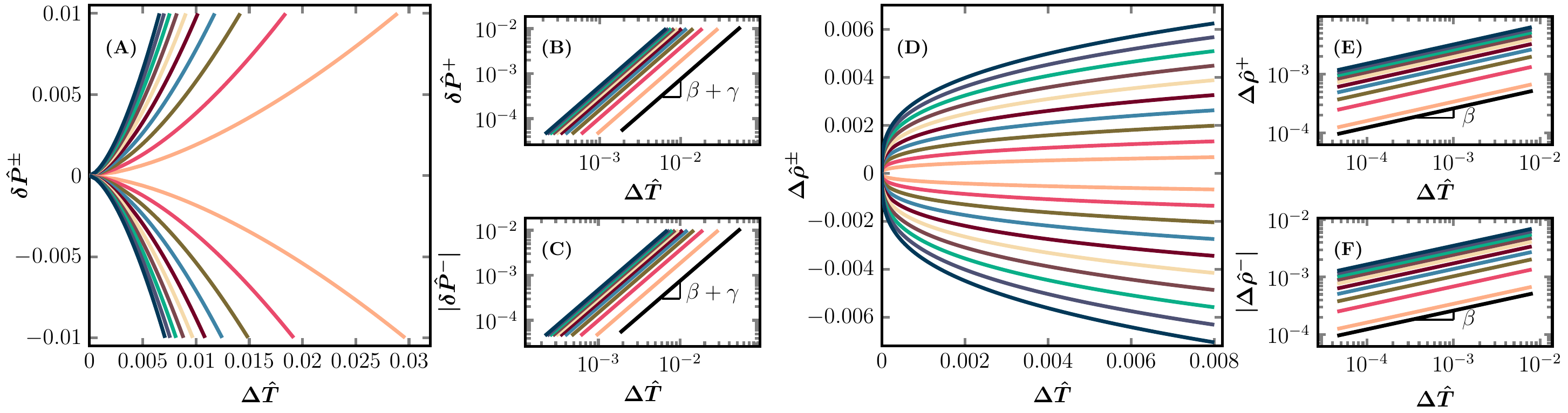}
  \caption{{\bf Thermodynamic crossover lines $L^{\pm}$ obtained from the linear scaling theory for asymmetric models ($\varphi =82^{\circ}$).} (A) $\delta \hat{P}$ as a function of $\Delta \hat{T}$ along $L^{\pm}$, for   $a=0.02, 0.04, 0.06, 0.08, 0.1, 0.12, 0.14, 0.16, 0.18, 0.2$ (from right to left) and $k=a$. 
  %\YJ{Figure legend way too small.}
  The same data are plotted in log-log scales in (B) and (C) for $L^+$ and $L^-$ respectively, confirming the scaling Eq.~(\ref{eq:scaling_P}).
  %some different parameters $a$ varies from 0.02 to 0.2 in $\delta \hat{P}-\Delta \hat{T}$ phase diagram, and their double logarithmic curves are shown in  (B),(C), which are parallel to $y=(\beta+\gamma)x$. 
  (D)  $\Delta \hat{\rho}$ as a function of $\Delta \hat{T}$ along $L^{\pm}$.   The same data are plotted in log-log scales in (E) and (F) for $L^+$ and $L^-$ respectively, confirming the scaling Eq.~(\ref{eq:scaling_rho}).
  %$L^{\pm}$ in $\Delta \hat{\rho}-\Delta \hat{T}$ phase diagram. (E),(F) shows their double logarithmic form, where the curves are parallel to $y=\beta x$. 
  The meaning of  colors are the same in all panels.
  %\YJ{Make the notations consistent with Fig. 3, (A) $\delta \hat{P}^{\pm}$, (B) $\delta \hat{P}^+$ (C) $|\delta \hat{P}^-|$, the same for D-F, same for Fig. S10. }
  }
  \label{fig:linear_scaling}
\end{figure}

When the slope of the coexistence line is not zero, i.e.,  $\varphi \neq 0$, it is difficult to obtain an analytical expression of $L^{\pm}$. One has to compute $L^{\pm}$ numerically following the procedure described in Sec.~\ref{sec:vdw}.
Without  loss of generality, we use the following setup: the critical exponents are taken from the 3D Ising model universal class, $\beta \simeq 0.3265$ and $\gamma \simeq 1.237$~\cite{guida1998critical} (other exponents can be computed using the scaling relations $\alpha + 2\beta + \gamma = 2$ and $(\delta -1)\beta = \gamma$). The slope of the coexistence line is set to be $\varphi=82^{\circ}$, which is close to that of many real substrates (such as $\mathrm{CO_2, N_2O, CH_3F},$ etc.).The parameter $a$ varies from 
0.02 to 0.2. 
To reduce the number of independent parameters, we set $k/a = 1$ following previous studies, as suggested by the Monte Carlo simulations of the 3D Ising model with short range interactions~\cite{fuentevilla2006scaled,kim2003crossover}. 
%The parameter $k/a \approx 1$ according to a Monte Carlo simulation for a three-dimensional equivalent neighbor Ising model ~\cite{fuentevilla2006scaled,kim2003crossover}, therefore we simply set $k=a$ in order to reduce the number of parameters. } 
%To reduce the number of independent parameters, we set $k/a = 1$ following previous studies, as suggested by the Monte Carlo simulations of the three-dimensional Ising model with short range interactions~\cite{fuentevilla2006scaled,kim2003crossover}. 
The numerical solution of crossover lines $L^{\pm}$ using the linear scaling theory indeed confirms Eqs.~(\ref{eq:scaling_P}) and (\ref{eq:scaling_rho}), as shown in  Fig.~\ref{fig:linear_scaling}. 
We emphasize that these universal scalings are independent of the specific choice of parameters. 

%confirm  the scalings Eqs.~(\ref{eq:scaling_rho}) and (\ref{eq:scaling_P}). 

\begin{figure}[!htbp]
  \centering
  \includegraphics[width=\linewidth]{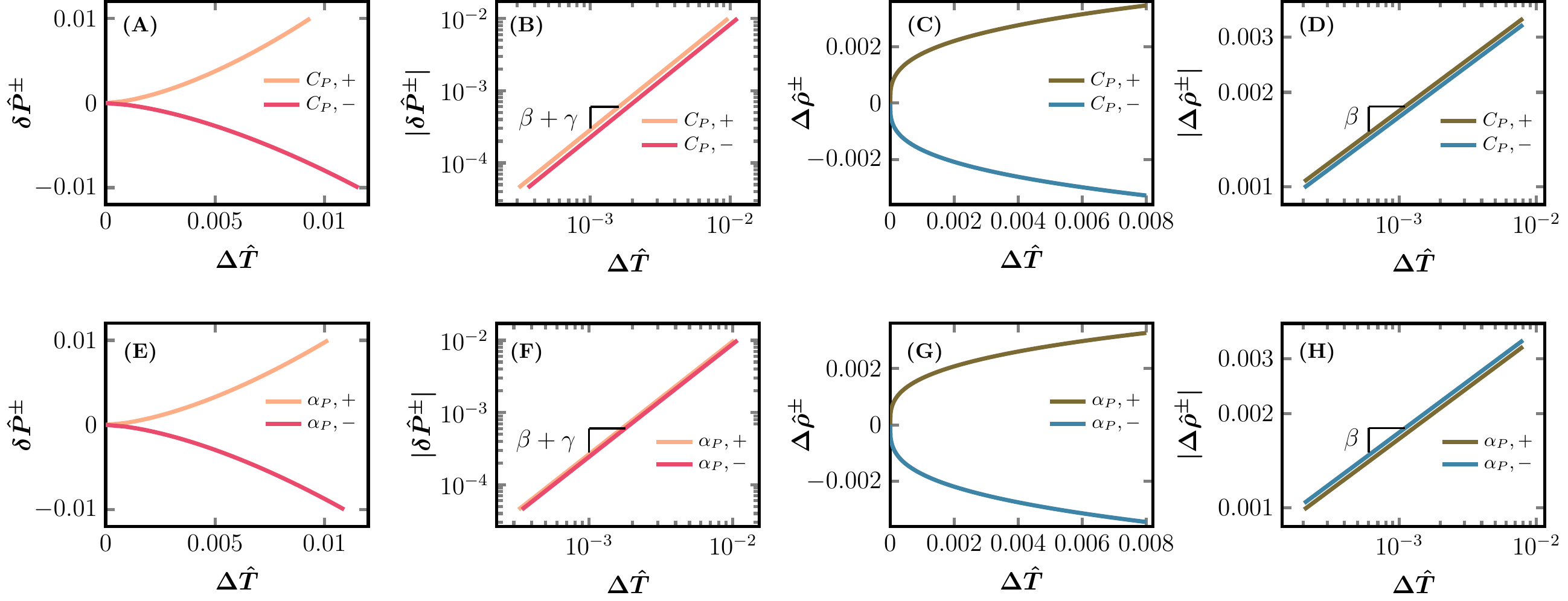}
  \caption{{\bf Thermodynamic crossover lines $L^{\pm}$ defined based on other response functions, obtained from the linear scaling theory for an asymmetric model ($\varphi =82^{\circ}$, $a=k=0.1$).}  % {\bf \blue{Other response functions for asymmetric models.}} 
  (A) $\delta \hat{P}$ as a function of $\Delta \hat{T}$ along $L^{\pm}$ defined according to  extreme values of the isobaric specific heat $C_{P}$; (B) same data in log-log scales. (C) $\Delta \hat{\rho}$ as a function of $\Delta \hat{T}$ along $L^{\pm}$ defined according to $C_{P}$; (D) same data in log-log scales. 
  (E-H) Similar to (A-D), but with $L^{\pm}$ defined according to  extreme values of the isobaric thermal expansion $\alpha_{P}$.
 In all case, the scalings Eqs.~(\ref{eq:scaling_P}) and (\ref{eq:scaling_rho}) are satisfied.  
 %\YJ{Replace $C_p$ by $C_P$, $ap$ by $\alpha_P$.}
 %\YJ{Make the order of panels consistent with the description. Show the scalings in the plots. keep $\Delta \hat{\rho}$ and $\Delta \hat{\rho}$ consistent in the paper, although they are identical (maybe $\Delta \hat{\rho}$ is better) .}
 % and (C) show extreme value lines of isobaric specific heat and isobaric thermal expansion in $\delta \hat{P}-\Delta \hat{T}$ phase diagram, and (E), (G) show them in $\Delta \hat{\rho}-\Delta \hat{T}$ phase diagram. (B), (D), (F) and (H) show double logarithmic form of curves in their left panels respectively.
%  \YJ{add lines to show the scaling as in fig. s7B}}
  }
  \label{fig:scaling_Cpap}
\end{figure}

\begin{figure}[!htbp]
  \centering
  \includegraphics[width=0.85 \linewidth]{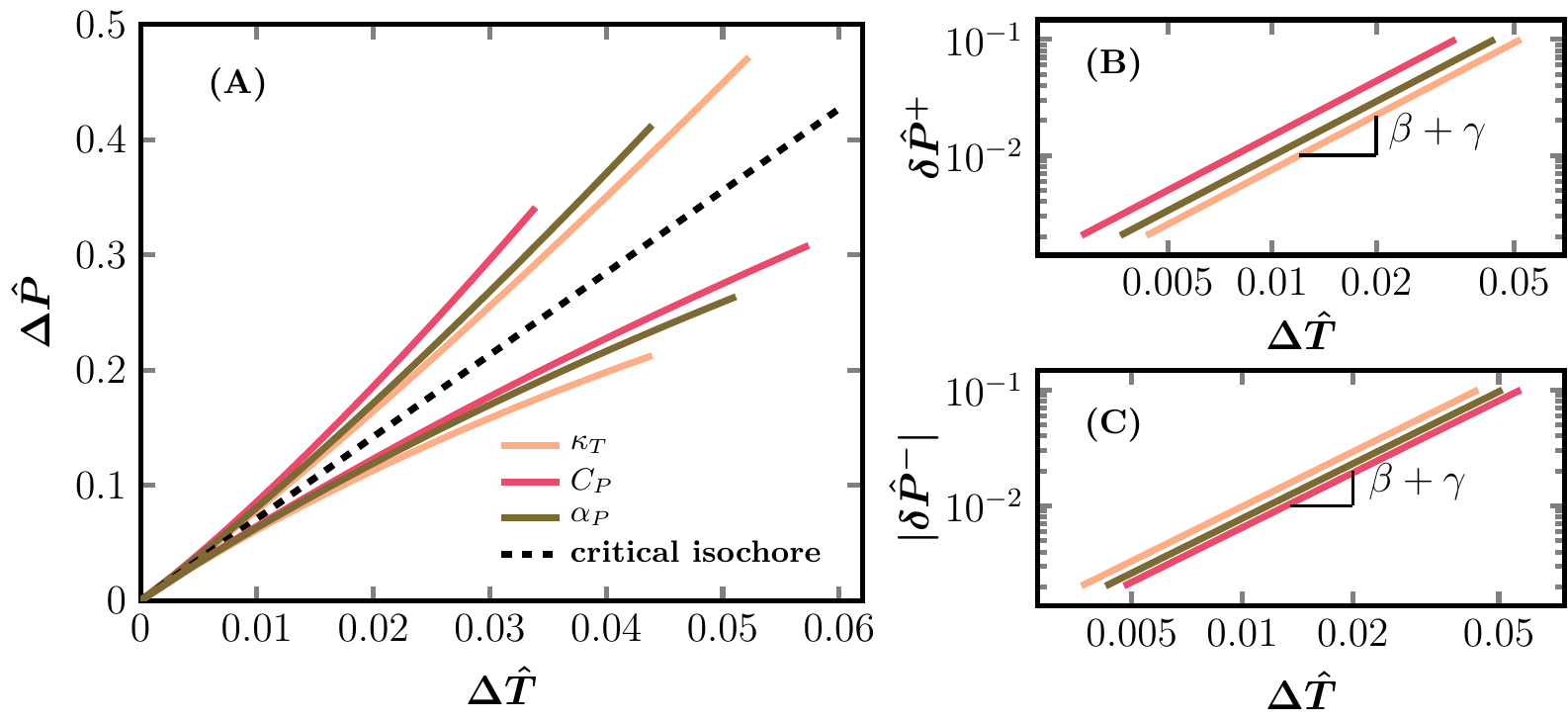}
  \caption{{ \bf Comparison of $L^{\pm}$  defined according to three different thermodynamic response functions in the linear scaling theory.} Results are obtained for $\varphi =82^{\circ}$ and $a=k=0.1$. Same data are plotted in log-log scales in (B) and (C).
  %\YJ{(C) $|\delta \hat{P}^-|$}
  %\YJ{replace $\kappa_T$ by $\kappa_T$; same for others (check other figures); replace 'isochore' by 'critical isochore' (check other figures). }
%   in original $\Delta \hat{P}-\Delta \hat{T}$ phase diagram.} Lines with 3 different colors represent 3 response functions $\kappa_T$, $C_P$ and $\alpha_P$, and they gradually separate with each other when away from critical point.}
  }
  \label{fig:scaling_3functions}
\end{figure}

Interestingly, the scalings Eqs.~(\ref{eq:scaling_P}) and (\ref{eq:scaling_rho}) remain satisfied if one defines the supercritical crossover lines as the lines of extremums of another response function, such as the isobaric specific heat $C_P$ or the isobaric thermal expansion coefficient $\alpha_P$. Following a similar procedure as described in Sec.~\ref{sec:vdw}, such crossover lines can be also obtained numerically (see Fig.~\ref{fig:scaling_Cpap}).  Without loss of  generality, we set $\varphi=82^{\circ}$ and $a = k = 0.1$.  Our results demonstrate that the scalings, Eqs.~(\ref{eq:scaling_P}) and (\ref{eq:scaling_rho}), are independent of which particular response function is used in the definition of $L^{\pm}$. 
While the scalings are universal, Fig.~\ref{fig:scaling_3functions} shows that the supercritical crossover lines defined according to different response functions do not completely coincide, in the regime that is not very close to the critical point.

%the crossover lines extreme value line of other two response functions, isobaric specific heat $C_P$ and isobaric thermal expansion $\alpha_P$, also satisfy these two scalings when $\varphi \neq 0$. The calculation should be followed the procedure described in Sec.~\ref{sec:vdw}, too. Here we randomly selected an adjustable parameter $a = k = 0.1$ to show the results, as can be seen in Fig.~\ref{fig:scaling_Cpap}, where the slope of curves in $\log(\delta \hat{P})-\log(\Delta \hat{T})$ phase diagram are all $\beta+\gamma$, and in $\log(\Delta \hat{\rho})-\log(\Delta \hat{T})$ phase diagram are all $\beta$.
%\YJ{Four panels: (A-B)  add the data for the scaling of $\rho$; (C) Fig.~S3A; (D) S3A-inset (maybe separate $L^+$ and $L^-$ as in Fig. 4B-C?)}

\color{black}

\section{
Summary of the coefficients $A_P^\pm$ and $A_\rho^\pm$ in Fig.3.}

Table~\ref{table:coefficients} summarizes the fitting parameters $A_P^\pm$ and $A_\rho^\pm$  for eight different substrates in Fig.~3.

\begin{table*}[!htbp]
\centering
\caption{Summary of the coefficients $A_P^\pm$ and $A_\rho^\pm$ in Fig.3.}

\begin{tabular}{lrrrr}
\qquad substrates \qquad & $\qquad A_{P}^+ \qquad$ & $\qquad   A_{P}^-  \qquad$ & $\qquad   A_{\rho}^+  \qquad$ & $\qquad   A_{\rho}^-  \qquad$\\ 
\midrule
$\rm{Ar}$  &  17.31 & 10.22 & 0.83 & 0.97 \\
$\rm{C_3H_8}$  & 21.98 & 15.08 & 0.85 & 0.94 \\
$\rm{CO_2}$  &  23.39 & 13.74 & 0.88 & 0.99 \\
$\rm{H_2O}$  & 23.83 & 15.54 & 1.06 & 0.98 \\
$\rm{N}$  & 18.44 & 9.97 & 0.85 & 0.99 \\
$\rm{N_2O}$  & 22.20 & 13.50 & 0.85 & 0.98 \\
$\rm{Ne}$ & 16.58 & 9.99  & 0.86 & 0.99 \\
$\rm{O_2}$ & 17.99 & 10.59 & 0.83 & 0.98 \\
\bottomrule
\end{tabular}
\label{table:coefficients}
\end{table*}

\section{Additional data of $L^{\pm}$ lines of argon}

%Discuss here different fitting methods. Difference between using $\kappa_T$ and $\beta$, etc. 
\subsection{Examination of different  fitting methods}
\label{sec:fitting}

\begin{figure}[!htbp]
  \centering
 \includegraphics[width=0.9\linewidth]{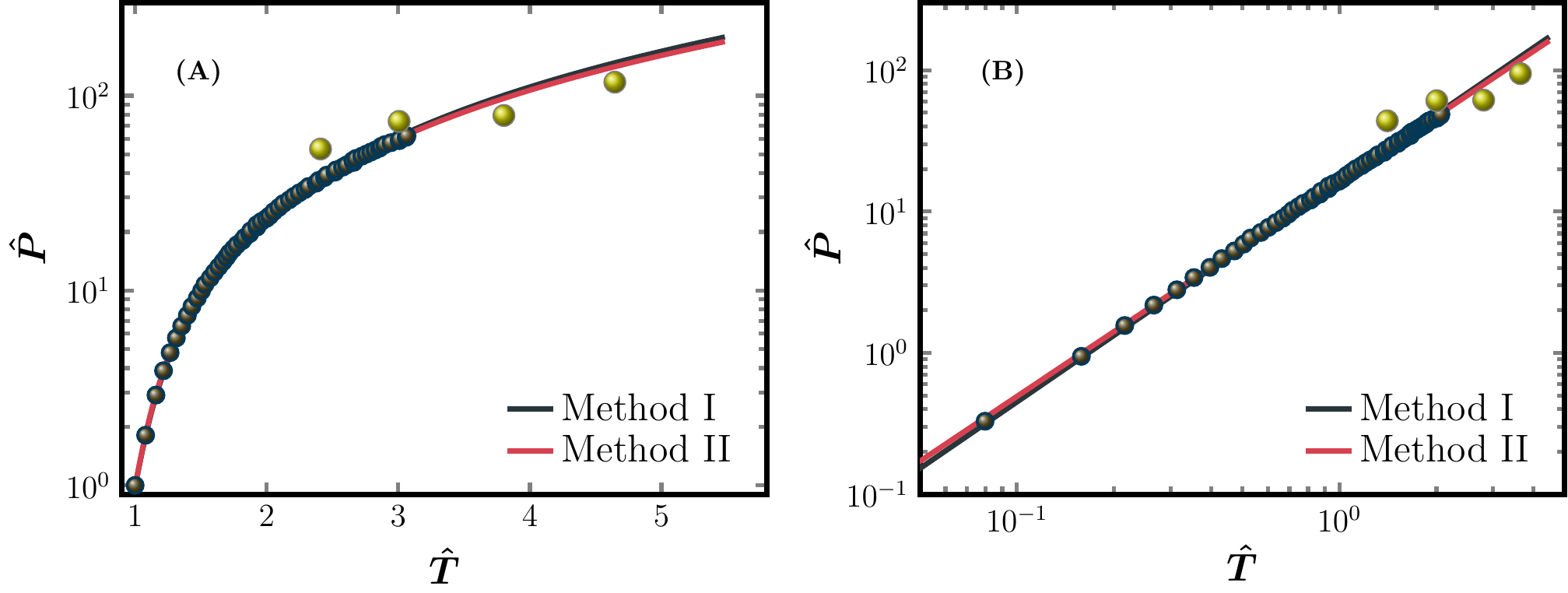}
  \caption{{\bf Comparison of the extrapolated $L^+$ line of supercritical argon using two
  different fitting methods.}
 % \YJ{Experimental data and $L^\pm$ points all look blue. I think here you can remove experimental points, $L^-$ line and critical isochore. Make the two colors of  solid lines as different as possible (for example black and red). Replace the legend by "Method I" and "Method II". }
  (A) 
  Cycles are obtained by locating maxima of $\kappa_T$ using the NIST data. Solid lines with different colors are obtained using two different fitting methods as described in the text. %\YJ{$\kappa_T$? check it everywhere}
  %belongs to argon dynamic crossover lines, which are approximately symmetric with respect to critical isochore (dashed line). Two different color lines correspond to two fitting equations, where 
  For the first method (I), the resulted fitting parameter is $A_P^+ = 16.49$. For the second method (II), the resulted fitting parameters are, $A_P^+ = 16.45$ and $\beta+\gamma=1.523$.
 Same data are plotted in log-log scales in (B).
 Yellow balls are dynamical crossovers estimated from the sound dispersion data (see Fig.~4).
 %\YJ{Add yellow balls in (B)}
  % $C_1^+=16.49$, $C_2^+=16.45$ and $\beta+\gamma=1.523$. (B) show the fitting on the log-log scale. The meaning of experimental data symbols are the same in maintext.}
  }
  \label{fig:fitting}
\end{figure}

By definition, $L^+$ is determined at the loci of maxima of $\kappa_T$ along paths parallel to the isochore. Away from the critical point, the maximum becomes difficult to locate due to the limited range of experimental EOSs. Thus one has to rely on numerical extrapolations to estimate $L^+$ at  pressures much higher than the critical pressure.
Based on the scaling obtained from the above theoretical analysis, we  fit the data of  $L^+$ near the critical point, which are directly determined from the maxima of $\kappa_T$ (circles in Fig. 4 and Fig.~\ref{fig:fitting}), to,
\beq
\delta \hat{P}^+ = A_P^+ \Delta \hat{T}^{\beta + \gamma}.
\label{eq:fit_Lp}
\eeq
 Then the fitting curves are extrapolated to the higher  pressure regime
(lines in  Fig. 4 and Fig.~\ref{fig:fitting}), in order to have a full comparison with the sound dispersion data. Two fitting methods are considered.  In the first method, we set  the 3D Ising model universal class critical exponents  $\beta+\gamma=0.326+1.237 = 1.563$, and treat the coefficient $A_P^+$ as a fitting parameter. In the second method,  $\beta + \gamma$ and $A_P^+$ are treated as  two independent fitting parameters. As shown in Fig.~\ref{fig:fitting}, the two methods give nearly identical extrapolations  in 
the interested 
region of the phase diagram.

\subsection{$L^{\pm}$ lines  defined by different response functions}
%\YJ{Make a figure of $L^+$ line defined according to four response functions, $\kappa_T$, $\beta_T$, $C_P$, $\alpha_P$.  Remove isochore and $L^-$. I think even the experimental data can be removed.}

%\blue{In gas-liquid phase transition, the compressibility is usually defined as $\beta_T = \frac{1}{\rho}\left(\frac{\partial \rho}{\partial P}\right)_T$, where the coefficient $\frac{1}{\rho}$ ensures $\beta_T$ an intensive quantity. We can also determine the crossover line by this general definition. In the same way as in the main text, we find the peak $\hat{T}_{\rm max}^{+}(\delta \hat{P})$ of $\beta_T$ for a fixed $\delta \hat{P}$, and they connected to another crossover line. We use Eq.~(\ref{eq:fit_Lp}) to fit the data using the 3D Ising model universal class critical exponents  as shown in Fig.~\ref{fig:kappacompare}, where we can see the data deviate from the scaling relationship at a much faster rate than $L^+$ calculated by $\kappa_T = \left(\frac{\partial\hat{\rho}}{\partial\hat{P}}\right)_{\hat{T}}$, and they don't accord with the experiment as well as $L^+$.}

\begin{figure}[!htbp]
  \centering
  \includegraphics[width=0.8 \linewidth]{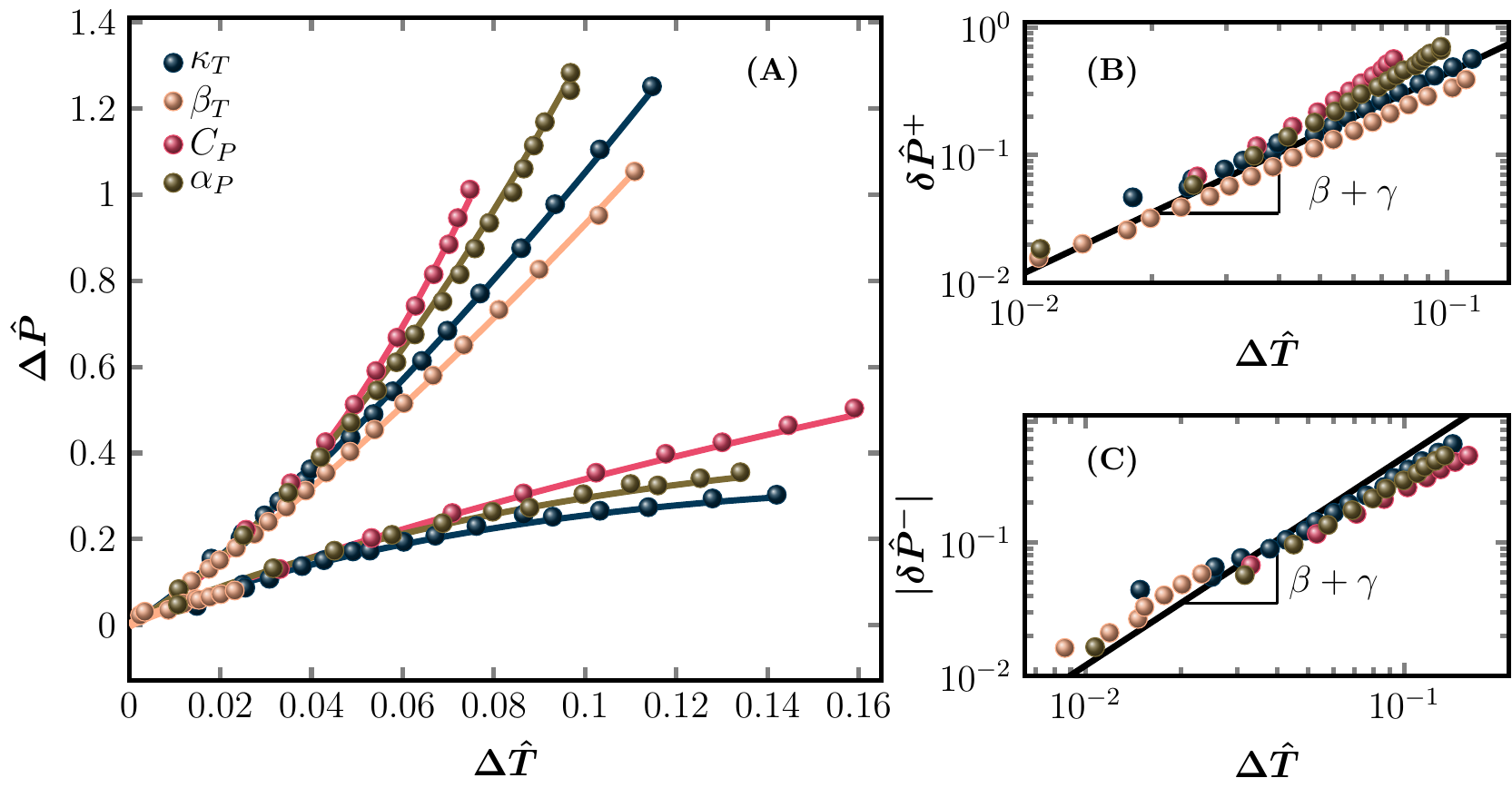}
  \caption{ {\bf Comparison of the $L^{\pm}$ lines defined according to four different thermodynamic response functions of argon.} (A)
  $L^{\pm}$ lines obtained from the maxima of four 
 different response functions,  $\kappa_T = \left(\frac{\partial\hat{\rho}}{\partial\hat{P}}\right)_{\hat{T}}$, $\beta_T = \frac{1}{\hat{\rho}}\left(\frac{\partial \hat{\rho}}{\partial \hat{P}}\right)_{\hat{T}}$, $C_P = \hat{T} \left(\frac{\partial \hat{S}}{\partial \hat{T}}\right)_{\hat{P}}$, and $\alpha_P = -\frac{1}{\hat{\rho}}\left(\frac{\partial \hat{\rho}}{\partial \hat{T}}\right)_{\hat{P}}$.
 %  following the same steps described in text. %Lines are fitting results. 
   (B) and (C) are the same data in log-log scales for $L^+$ and $L^-$ respectively. Solid lines represent the scaling $\delta \hat{P}^{\pm} \sim \Delta \hat{T}^{\beta+\gamma}$ with $\beta+\gamma = 1.563$. 
   %\YJ{(A) unify with Fig. S11, $\Delta \hat{P}$ vs $\Delta \hat{T}$. (C) $|\delta \hat{P}^-|$}
   %\YJ{In B and C, change $\delta \hat{P} $ to $\delta \hat{P}^+$ ...}
  }
  \label{fig:Ar_4functions}
\end{figure}

Here we check that if $L^{\pm}$ lines can be defined using other response functions (see Fig.~\ref{fig:Ar_4functions}). 
As shown in Fig.~\ref{fig:Ar_4functions}A, even though these lines do not completely coincide, the discrepancies reduce approaching the critical point.
Further more, they follow the same scaling (Fig.~\ref{fig:Ar_4functions}B and C), consistent with the theoretical results calculated using the linear scaling theory (see Fig.~\ref{fig:scaling_3functions}).
%\YJ{in Fig. S10, also add two log-log plots}
%Even though there depart from each other when away from critical point, all of them satisfy the scaling-law. However, they deviate from the scale law with different rates, as can be seen in Fig.~\ref{fig:Ar_4functions}(B),(C).}
Unlike phase transition lines where all response functions are singular, the thermodynamical crossover lines $L^{\pm}$ defined by different response functions are not unique, except for the universal scalings. 
When the crossover line is compared to experimental data, one should choose a proper response function suitable for the given experiment. For example, in the inelastic X-ray scattering experiment, the measured data of sound speed and dispersion are essentially related to  density fluctuations~\cite{gorelli2013dynamics, simeoni2010widom};  thus is is reasonable to choose  $\kappa_T$ to define $L^{\pm}$ in order to compare with the  data (see Fig.~4 and Sec.~\ref{sec:argon_sound}). %\YJ{Plot the three kinds of crossover lines in one figure $\Delta \hat{P}$ -$\Delta \hat{T}$ (not rotated), similar to Fig. 1a of Xu PRL paper.}

\section{Reestimation of  dynamical crossovers in supercritical  argon from sound dispersion  data}
\label{sec:argon_sound}
\begin{figure}[!htbp]
  \centering
 \includegraphics[width=\linewidth]{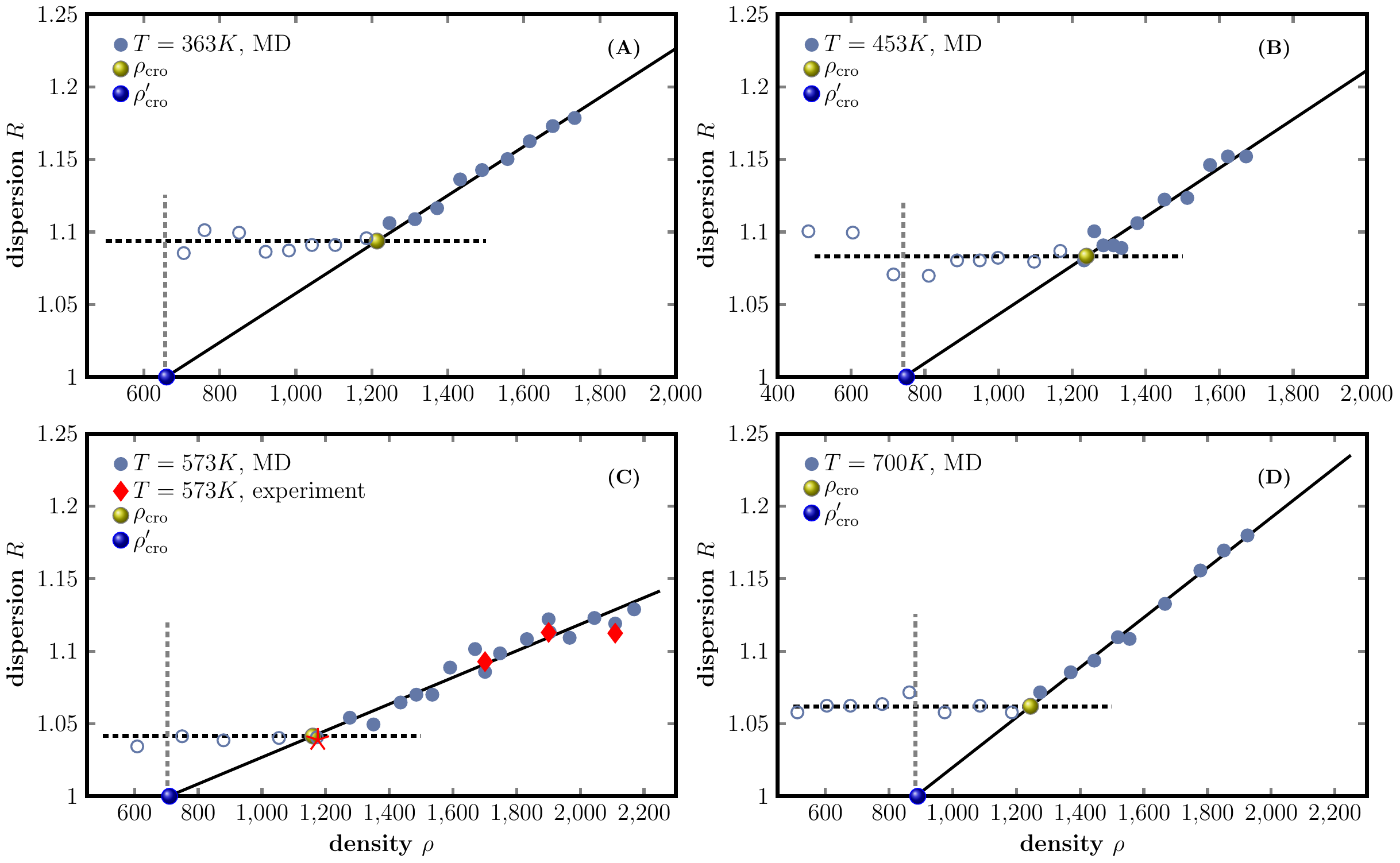}
  \caption{ {\bf Re-estimation of dynamical crossovers in supercritical argon, based on the sound dispersion data collected from~\cite{gorelli2013dynamics, simeoni2010widom}.}
  % Redistribution of sound dispersion MD simulation data.} 
  %\YJ{check T vs $T$, R vs $R$; check this everywhere}
 % \YJ{In the figure, change "Dispersion" to "dispersion $R$", "Density" to "density $\rho$".}
  Sound dispersion $R$ as a function of density $\rho$ at four different temperatures. 
  Open and filled circles are data points obtained from molecular dynamics (MD) simulations  in ~\cite{gorelli2013dynamics, simeoni2010widom}.
  Open circles are fitted to a constant (horizontal dashed line); filled circles are fitted to a linear function (solid line).
%Hollow dots corresponds to those tends to be a constant, whose value are indicated by dashed line with the corresponding color, and solid dots represent positive dispersion that increase with density, and their linear fitting lines are black lines. 
  The intersection of dashed and solid lines defines the crossover density $\rho_{\rm cro}$.
  %\YJ{In (A) and (C), it looks like the rightmost open point should be filled. }
  The intersection between the solid line and the $R=1$ axis defines $\rho'_{\rm cro}$, according to Ref.~\cite{gorelli2013dynamics}.
  %points of these two lines are crossover point in our definition. 
  Red diamonds in (C) represent experimental data and the red star represents the crossover estimated in ~\cite{simeoni2010widom}.
  %\YJ{Can't see the red star, make it larger or thicker. B and C: two $\rho'$; (C) update or remove the reference [6] }
  %$\rho^{\prime}_{cro}$ can be defined when black line tends to 1 at y-axis, in other words the intersection points with gray dashed line.}
  %\YJ{D: keep the same range of y-axis: from 1 to 1.25.}
  %\YJ{Indicate $\rho_{\rm cro}$ and $\rho'_{\rm cro}$ in the figure by points, and indicate of them as $\rho_{\rm cro}$ and $\rho'_{\rm cro}$ in the legend (don't indicate the dashed lines as $\rho_{\rm cro}$ and $\rho'_{\rm cro}$).  Keep the symbol of $\rho_{\rm cro}$  consistent with that in Fig. 3. Remove the vertical dashed lines. In (C), remove  "experimental crossover" (the crossover was estimated using both experimental and simulation data in [6]). Use same symbols/colors of MD points at all $T$ and keep them consistent with Fig. 3. }
  }
  \label{fig:sound_dispersion_data}
\end{figure}

We collect experimental and simulation   sound dispersion data of supercritical argon from the literature~\cite{gorelli2013dynamics, simeoni2010widom}, and re-analyze them to obtain  supercritical crossovers.
The data are reported at four temperatures, 363 K ($2.4 T_{\rm c}$), 453 K ($3 T_{\rm c}$), 573 K ($3.8 T_{\rm c}$) and 700 K ($4.6 T_{\rm c}$).
Several data points at $T=573K$ are obtained from 
inelastic X-ray scattering experiments in~\cite{simeoni2010widom} (see Fig.~\ref{fig:sound_dispersion_data}); %\YJ{mark experimental points with a different symbol or color} 
other data points are obtained from molecular dynamics (MD) simulations in~\cite{gorelli2013dynamics, simeoni2010widom}. In Fig.~\ref{fig:sound_dispersion_data}, we plot the positive dispersion ratio $R = c_{\rm L}^{\rm max} / c_{\rm s}$ as a function of density $\rho$, where  $c_{\rm L}^{\rm max}$ is the maximum of $c_{\rm L} (Q)$,  $c_{\rm s} = c_{\rm L} (Q \to 0) $   is the adiabatic sound velocity, and $c_{\rm L} (Q)$  is the 
longitudinal sound velocity as a function of the wave number $Q$. A zero value of $R$ means the absence of dispersion. 

In Ref.~\cite{simeoni2010widom},  a supercritical dynamical crossover is defined as the separation between a gas-like regime, where $R \approx 1.04$ is a small constant, and a liquid-like regime, where $R$ increases with the pressure $P$, under a fixed temperature $T=573$~K. We follow this definition to evaluate the crossover points at other temperatures. Specifically, we fit $R(\rho)$   to a linear function at large $\rho$, and  a constant at small $\rho$. The intersection of the two lines determines the crossover density $\rho_{\rm cro}$. Such a method is in fact commonly used in the evaluation of other dynamical crossovers, e.g., the evaluation of liquid-glass transition temperature $T_{\rm g}$ from heat capacity data. 
Once $\rho_{\rm cro}(T)$ is available, $P_{\rm cro}(T)$ can be computed according to  the EOSs provided by the NIST database. Note that our estimated crossover at $T=573$ K is fully consistent with the value reported in ~\cite{simeoni2010widom}. 
%\YJ{add the crossover point in ~\cite{simeoni2010widom} to Fig. S5}

In Ref.~\cite{gorelli2013dynamics}, the authors  use another definition to evaluate the supercritical crossovers. The linear fitting of $R(\rho)$ at large $\rho$ is extended to lower $\rho$; 
the intersection  between this linear extrapolation and the horizontal line $R=1$ defines $\rho_{\rm cro}'$. 
As can be seen in Fig.~\ref{fig:sound_dispersion_data},  $\rho_{\rm cro}'$ obtained in this way
is already in the constant $R(\rho)$ regime, and therefore might 
underestimate the crossover density.
%, since it is already in the constant $R(\rho)$ regime. In other words, $\rho_{\rm cro}'$ cannot effectively separates two kinds of behavior: constant $R(\rho)$ and  increasing $R(\rho)$. 
Note that the criterion  $R=1$ (i.e., the complete absence of dispersion) is assumed for the gas or gas-like regime in Ref.~\cite{gorelli2013dynamics}. However,  according to our phase diagram (Fig.~4), the constant $R(\rho)$ data are in the liquid-gas indistinguishable regime, rather than in the gas-like regime. 
There is no particular reason to assume that the dispersion is completely absent in the indistinguishable regime: the differences on the acoustic properties of the indistinguishable and gas-like regimes remain to be explored in future studies. 

Based on above discussions, we find that  $\rho_{\rm cro}$ is a more reliable and accurate estimation of the supercritical  crossover density than $\rho_{\rm cro}'$.  Thus the former is adopted in this study.

%For the above reasons, we do not use the definition of $\rho_{\rm cro}'$ in our study.
%\YJ{separate Fig. S5 into four panels. Include both definitions $\rho_{\rm cro}$ and $\rho_{\rm cro}'$.}

%In Fig.~\ref{fig:experiment_L+}, the original MD simulation data for the sound dispersion in the supercritical region of Ar are from Gorelli et al~\cite{gorelli2013dynamics}. Their viewpoint is that dynamic crossover occurs when positive sound dispersion is completely disappeared, which is different from ~\cite{simeoni2010widom}. The latter thinks that dynamic crossover occurs at where the dispersion tends to be a constant. This point is more reasonable in our view. Therefore, we redefine the dynamic crossover of MD simulation data for the sound dispersion using the latter method, as shown in Fig.~\ref{fig:sound_dispersion_data}. The solid dots satisfy the equation $y=k x+b$, while hollow dots satisfy the equation $y=a$. The intersection point of these two equations is where dynamic crossover occurs at each temperature.

\section{Comparison of different supercritical crossover lines of argon in the vicinity of the critical point}
\begin{figure*}[!htbp]
  \centering
  \includegraphics[width=0.68\linewidth]{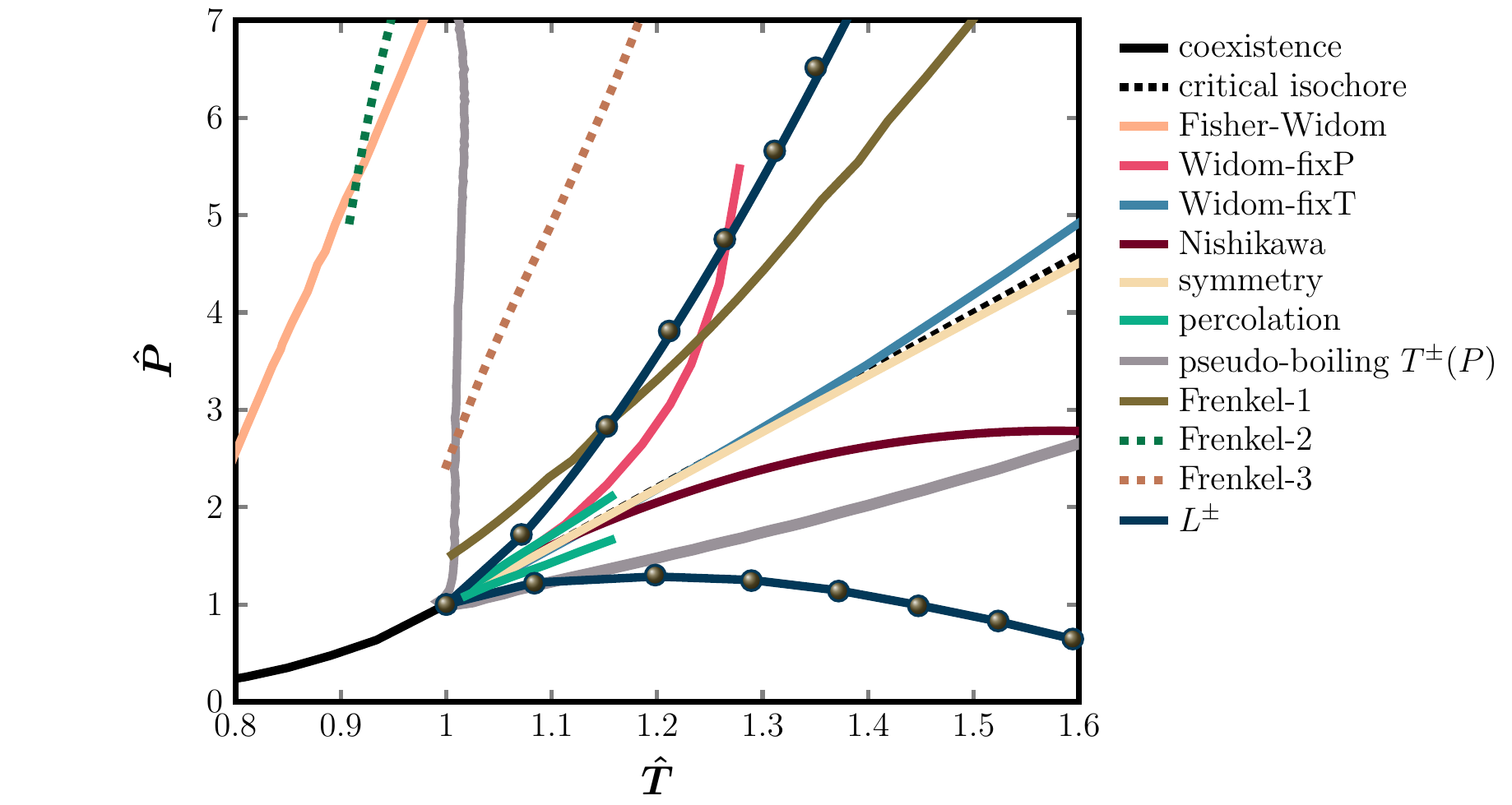}
  \caption{{\bf Different supercritical crossover lines of argon in the vicinity of the critical point.} This is an enlarged view of Fig.~4(A), and all symbols and colors represent the same meanings as in Fig.~4(A). %\YJ{$L^\pm$, also Fig. 4}
  %\YJ{change 'Symmetry' to 'symmetry'; $T/Tc$ to $\hat{T}$...replace $L^{+-}$ by $L^{\pm}$}
  %\YJ{The width of all lines can be increased. The size of points can be increased. This applies to most figures. Fig. S9 is a good example. In most figures, the font size can be increased. Readability is important.}
  }
  \label{fig:experiment_enlarged}
\end{figure*}

%\YJ{Move inset of Fig. 3A to SI.}
In Fig.~\ref{fig:experiment_enlarged}, we compare $L^{\pm}$ lines with other crossover lines of supercritical argon near the critical point. This is an enlarged view of Fig.~4(A) in the vicinity of the critical point, presented in linear scales. 

\section{~Nishikawa line of argon}

\begin{figure}[!htbp]
  \centering
 \includegraphics[width=0.6\linewidth]{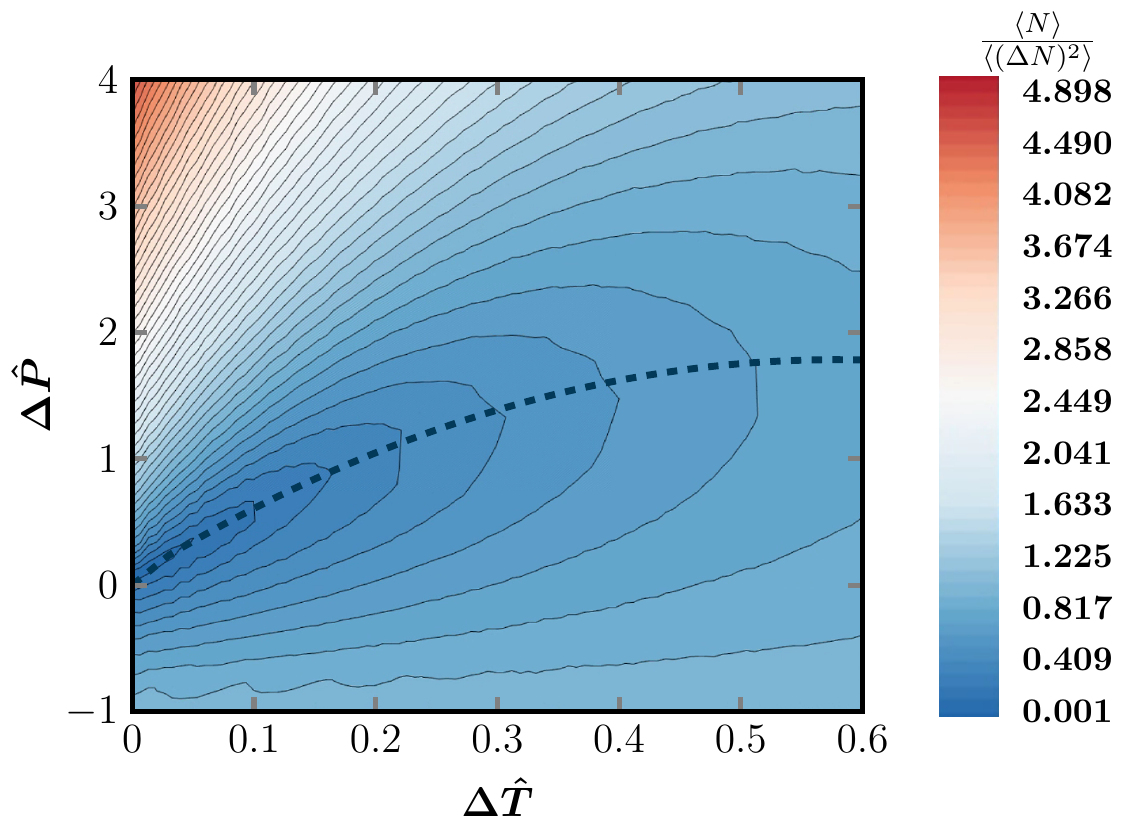}
  \caption{{\bf Colormap of the density fluctuation $\frac{\langle(\Delta N)^2\rangle}{\langle N \rangle}$ in the $P$-$T$ phase diagram.} Thin solid lines are constant density fluctuation lines. The dashed black line is the ridge that defines the Nishikawa line.
  %\YJ{I think $1/\kappa_T$ is not identical to the inverse of $\frac{\langle(\Delta N^2)\rangle}{\langle N \rangle}$, according to Eq. (S34), there is a factor of T difference}
  %\YJ{colormap of $1/\kappa_T$? add color bar, indicate the third axis on the color bar, $1/\kappa_T$ }
  }
  \label{fig:Nishikawa_line}
\end{figure}

\begin{figure*}[!htbp]
  \centering
  \includegraphics[width=0.9\linewidth]{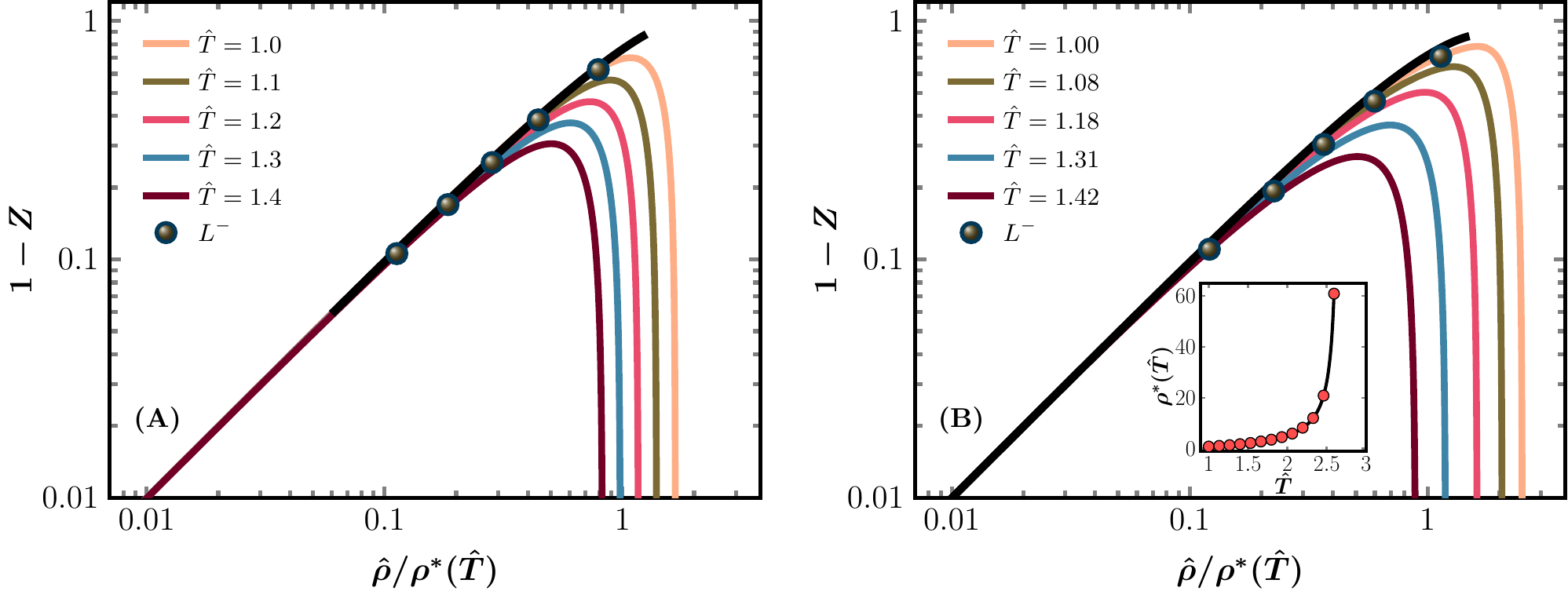}
  \caption{{\bf Validations of the $L^-$ line by the equation of states for the compressibility factor.} 
  (A) Van der Waals EOS. Colored lines are isotherms and the black line is Eq.~(\ref{eq:Zgas}) with $\rho^*(\hat{T}) = 24 \hat{T}/(27 - 8 \hat{T})$ and $\mathcal{F}(x) =  -x + 0.24 x^2$.
  (B) NIST data of argon. %Colored lines are isother. 
  %Validation of $L^-$ with experimental data through the EOS of compressibility factor $z$, which can be a single EOS of rescaled density $\rho/\rho^*(\hat{T})$ in the gas-like regime. 
  Colored lines are isotherms and the black line is the fitted gas-like EOS  Eq.~(\ref{eq:Zgas}) with $\mathcal{F}(x) = -x + 0.28 x^2$.
  The points of $L^-$ line are determined independently in Fig.~2, which coincide with the deviation points as shown in the plot.
  %and when  away from the gas-like regime, those different curves (represents a few different temperatures) deviate from the EOS, starting from points (dark blue) belongs to the $L^-$ line. 
  The inset shows $\rho^*(\hat{T})$ as a function of $\hat{T}$, where the solid line represents $\rho^*(\hat{T}) =  1/(-0.57+1.51/\hat{T}) $ obtained from fitting.%\YJ{$\hat{T}$ in legend? Add (A) and (B) in the figure} 
  }
  \label{fig:L-}
\end{figure*}

Because the Nishikawa line of argon is not available in the literature, here we estimate it according to its definition. 
The Nishikawa line is determined by the ``ridge'' of density fluctuations $\frac{\langle(\Delta N)^2\rangle}{\langle N \rangle}$ on the $P-T$ phase diagram.
%\YJ{should be  $\frac{\langle(\Delta N)^2\rangle}{\langle N \rangle}$?}
The density fluctuation  can be computed from EOSs using the following formula~\cite{nishikawa1998fluid}:
\beq
\frac{\langle(\Delta N)^2\rangle}{\langle N \rangle} %= \frac{N}{V} \kappa_T k_B T 
=R T \rho \beta_{T},
%= R T \frac{\partial \rho}{\partial P},
\label{eq:density_fluctuation}
\eeq
where 
%$K_B$ is Boltzmann constant, 
$R$ is universal gas constant.
%,
%$N$ is the number of particles, 
%and $\beta_T$ is isothermal compressibility, 
%\beq
%\beta_T = -\frac{1}{V} \left(\frac{\partial V}{\partial P}\right)_T = \frac{1}{\rho} \left(\frac{\partial \rho}{\partial P}\right)_T.
%\label{eq:isothermal compressibility}
%\eeq
%The dimension of $R$, $T$, molar density $\rho$ and $P$ should be $J/(mol\cdot K)$, $K$, $mol/m^3$, $N/m^2$ respectively to ensure that Eq.~(\ref{eq:density_fluctuation}) is dimensionless. 
The data of argon isotherms are collected from the NIST database.
%we calculate its density fluctuations. 
Figure~\ref{fig:Nishikawa_line} shows the colormap and contour lines of density fluctuations on $P-T$ phase diagram, where the Nishikawa line is determined as the ridge of contour lines. The  Nishikawa line plotted in Fig.~4(A) is obtained in this way.

\section{~Validation of $L^-$ line by the equation of states for the compressibility factor}

We show that the $L^-$ line can be validated  by the behavior of EOS that relates compressibility factor $Z=PV/nRT$ to $\hat{\rho}$ and $\hat{T}$, where $n$ is the number of moles.
Specifically, the supercritical gas-like  states at different $\hat{T}$ obey a single EOS,
\beq
Z_{\rm gas}(\hat{\rho}, \hat{T}) - 1 = \mathcal{F} \left[\hat{\rho}/ \rho^*(\hat{T}) \right],
\label{eq:Zgas}
\eeq
and that the $L^-$ line coincides with the points where the actual EOS starts to deviate from this gas form.
%Eq.~(\ref{eq:Zgas}). 
%\YJ{check and unify `van der Waals' vs `van der Waals'; then make the notation $Z_{\rm VdW}$ consistent with it}
We first derive Eq.~(\ref{eq:Zgas}) from the van der Waals equation Eq.~(\ref{eq:van der Waals equation}), which can be rewritten as,
\beq
Z_{\rm vdW}(\hat{\rho}, \hat{T})   = \frac{1}{1-\frac{1}{3}\hat{\rho}} - \frac{9 \hat{\rho}}{8\hat{T}}.
\label{eq:vanderWaals2}
\eeq
In the dilute gas limit $\rho \to 0$, it can be expanded as,
\beq
Z_{\rm vdW}(\hat{\rho}, \hat{T}) 
 \approx 1 + \frac{\hat{\rho} }{3 } - \frac{9 \hat{\rho}}{8\hat{T}}.
\label{eq:vanderWaals3}
\eeq
Equation~(\ref{eq:vanderWaals3}) satisfies the form of Eq.~(\ref{eq:Zgas}),
with $\rho^*(\hat{T}) = 24 \hat{T}/(27 - 8 \hat{T})$ and $\mathcal{F}(x) \approx  -x$. For larger $\hat{\rho}$, Eq.~(\ref{eq:Zgas}) is still obeyed, with a modified $\mathcal{F}(x) \approx  -x + 0.24 x^2$ that can be obtained from fitting (see Fig.~\ref{fig:L-}A). 
The modification  to the dilute-gas EOS with $\mathcal{F}(x) \approx  -x$ implies corrections 
required for high-density gas-like states.
At even larger $\hat{\rho}$, Eq.~(\ref{eq:Zgas}) can no longer be satisfied  by modifying $\mathcal{F}(x)$. For a given $\hat{T}$, the point where the original van der Waals EOS Eq.~(\ref{eq:vanderWaals2}) deviates from the gas form Eq.~(\ref{eq:Zgas}) coincides with the $L^-$ line, as shown in  Fig.~\ref{fig:L-}(A). This result is consistent with our definition of  $L^-$ line: below the line the system is in the gas-like state that can be described by a uniform EOS Eq.~(\ref{eq:Zgas}); above the line the system enters into the liquid-gas indistinguishable state and thus the gas-like EOS Eq.~(\ref{eq:Zgas}) does not hold anymore. 
%\YJ{add a figure for  van der Waals} 

We find that the NIST data of argon also follow Eq.~(\ref{eq:Zgas}) with $\rho^*(\hat{T}) =  1/(-0.57+1.51/\hat{T}) $ and $\mathcal{F}(x) \approx   -x + 0.28 x^2 $ in the gas-like regime, as shown in Fig.~\ref{fig:L-}(B). For each $\hat{T}$, the gas-like regime begins from the ideal gas limit ($\rho \to 0$) and terminates at a point where the  EOS deviates from Eq.~(\ref{eq:Zgas}).
%that is only suitable for the gas state. 
The deviation points can be considered as an independent determination of the $L^-$ line, coinciding with those defined in Fig.~2.

%(see Fig.~\ref{fig:experiment_L+}A and C for a comparison).

\section{~Comparison between 
density fluctuations estimated from NIST EOSs and those measured in scattering experiments}

\begin{figure*}[!htbp]
  \centering
  \includegraphics[width=0.9\linewidth]{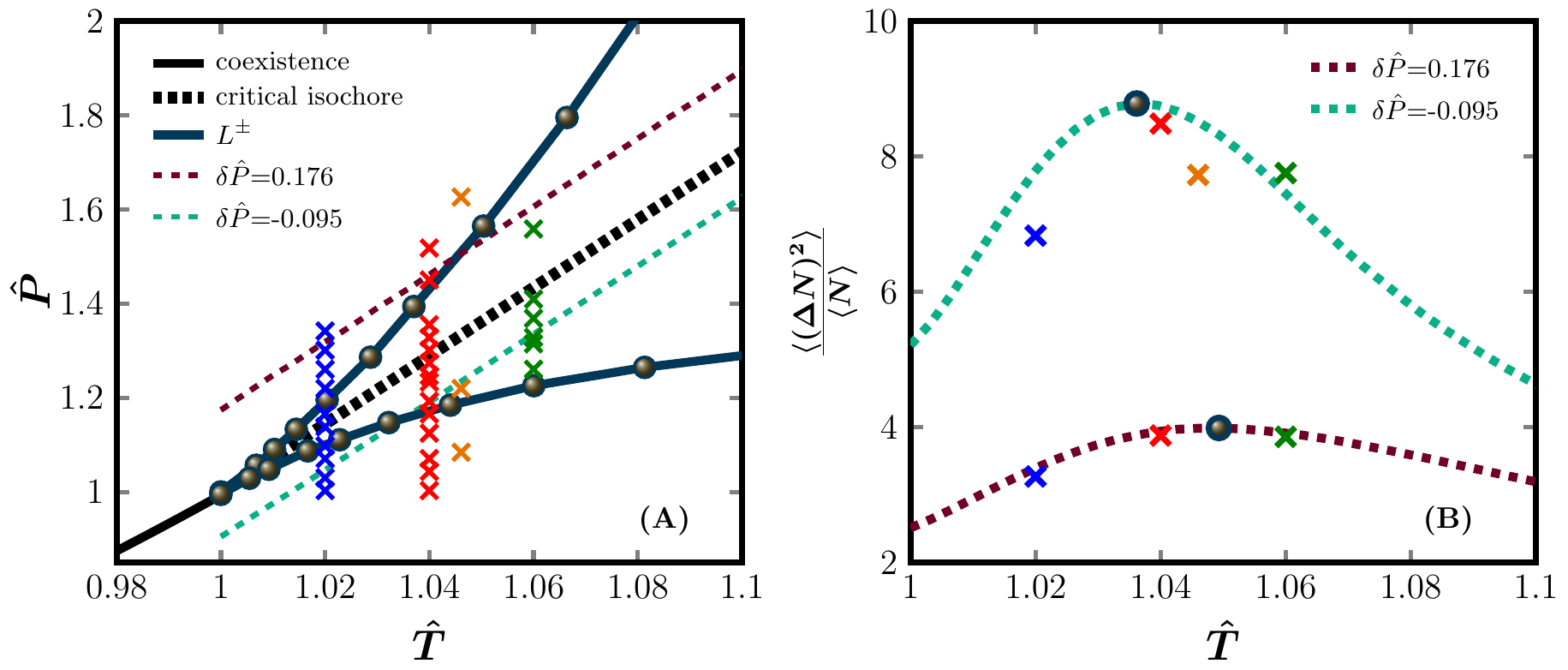}
  \caption{{\bf Comparison of density fluctuations derived from NIST EOSs with those measured in experiments.} (A) Enlarged $\hat{P}-\hat{T}$ phase diagram Fig.~4B of carbon dioxide $\rm CO_2$ near the critical point.
  Experimental data points (crosses) at four  different temperatures, $\hat{T}=1.02$ (blue), $\hat{T}=1.04$ (red), $\hat{T}=1.046$ (orange) and $\hat{T}=1.06$ (green), are collected from Refs.~\cite{nishikawa2003density, pipich2018densification}.
  For the comparison, we choose two example paths parallel to the critical isochore  at $\delta \hat{P}=0.176$ and $\delta \hat{P}=-0.095$. 
  In (B), we plot density fluctuations along the two paths obtained in two different ways.  
  Dashes lines represent density fluctuations obtained using Eq.~(\ref{eq:density_fluctuation2}) based on the NIST EOSs. 
  Crosses represent density fluctuations measured in previous experiments~\cite{nishikawa2003density, pipich2018densification}.
  The $L^+$ and $L^-$ crossover points (circles) correspond to the maxima of dashed lines. 
  %The orange points are SANS experimental data under $\hat{T}=1.46$, which are same with data in Fig.4(B). Two dashed lines are isochore paths when $\delta \hat{P}=0.176$(purple) and $\delta \hat{P}=-0.095$(green) respectively. (B) Density fluctuation $\frac{\langle(\Delta N)^2\rangle}{\langle N \rangle}$ as a function of $\hat{T}$, under $\delta \hat{P}=0.176$(purple) and $\delta \hat{P}=-0.095$(green). The points are the experimental data that these two paths pass through, the same colors with data in (A) reprsents the same temperature. 
  }
  \label{fig:DF_expdata}
\end{figure*}

In the main text (Figs.~2C and D), the susceptibility  $\kappa_T$ is obtained by taking numerical derivatives of the EOSs from the NIST database, and the  $L^+$ and $L^-$ lines are determined by the maxima of $\kappa_T$ along corresponding paths. The susceptibility  $\kappa_{T}$ is related to the density fluctuation $\frac{\langle(\Delta N)^2\rangle}{\langle N \rangle}$ by, 
\beq
\frac{\langle(\Delta N)^2\rangle}{\langle N \rangle} %= \frac{N}{V} \kappa_T k_B T 
=R T \kappa_{T}.
%= R T \frac{\partial \rho}{\partial P},
\label{eq:density_fluctuation2}
\eeq
The density fluctuation can be measured in scattering experiments, but the resolution is limited. To test the reliability of $L^+$ and $L^-$ lines determined in this study, 
here we collect the experimental carbon dioxide data from the literature and compare them with the density fluctuations calculated using NIST EOSs.
In Fig.~\ref{fig:DF_expdata}A,  the data points at $\hat{T}=1.02$, $\hat{T}=1.04$ and $\hat{T}=1.06$ were obtained by the small-angle x-ray scattering(SAXS) experiment in Ref.~\cite{nishikawa2003density}, and those at $\hat{T}=1.046$ were obtained by the small-angle neutron scattering(SANS) experiment~\cite{pipich2018densification}.
To locate $L^+$ and $L^-$ crossovers, we take two example paths parallel to the critical isochore, at $\delta \hat{P}=0.176$ and $\delta \hat{P}=-0.095$.
The density fluctuations derived from the NIST EOSs are plotted by dashed lines in Fig.~\ref{fig:DF_expdata}B, which are close to the experimental data points. Consequently, the maxima at $L^+$ and $L^-$ determined  from the NIST data are compatible with direct estimates using the experimental data within the resolution.

\end{document}